\documentclass[pre]{revtex4}
\usepackage{graphicx}
\usepackage{amssymb}
\usepackage{amsmath}
\usepackage{epstopdf}
\usepackage{array}
\usepackage{subfigure}
\usepackage{color}

\definecolor{gray}{rgb}{0.7,0.7,0.7}

\usepackage{supertabular}
\usepackage{hhline}
\newcommand\arraybslash{\let\\\@arraycr}
\makeatother
\newtheorem{finding}{Finding}
\newtheorem{definition}{Definition}

\usepackage{lscape}

\begin{document}

\title{Patterns of conductivity in excitable automata with updatable intervals of excitations}
\author{Andrew Adamatzky}
\affiliation{University of the West of England, Bristol, UK}
%\email{andrew.adamatzky@uwe.ac.uk}

\markboth{}{\tiny{Adamatzky A., Patterns of conductivity in excitable automata with updatable intervals of excitations. Phys. Rev. E 86, 056105 (2012) [16 pages] }}

\begin{abstract}
We define a cellular automaton where a resting cell excites if number of its excited neighbours belong to some specified interval and boundaries of the interval change depending on ratio of excited and refractory neighbours in the cell's neighbourhood. We calculate excitability of a cell as a number of possible neighbourhood configurations that
excite the resting cell. We call cells with maximal values of excitability conductive. In exhaustive search of functions of 
excitation interval updates we select functions which lead to formation of connected configurations of conductive cells. 
The functions discovered are used to design conductive, wire-like, pathways in initially non-conductive arrays of cells. 
We demonstrate that by positioning seeds of growing conductive pathways it is possible to implement a wide range 
of routing operations, including reflection of wires, stopping wires, formation of conductive bridges and generation of new wires in the result of collision. The findings presented may be applied in designing conductive circuits in excitable non-linear media, reaction-diffusion chemical systems, neural tissue and assembles of conductive polymers.
 \end{abstract}


\maketitle

\section{Introduction}

 Excitable cellular automata are well endowed tools for  studying complex phenomena of spatio-temporal 
physical, chemical and biological systems~\cite{ilachinski,chopard}, prototyping of chemical 
media ~\cite{gerhardt,markus_hess_1990},   reaction-diffusion computers~\cite{adamatzky_book_2005}, 
studying calcium wave  dynamics~\cite{yang}, and chemical turbulence~\cite{hartman}.

In a classical Greenberg-Hasting~\cite{greenberg-hasting} automaton model of excitation a cell takes 
three states --- reseting, excited and refractory. A resting cell becomes excited if number of excited neighbours
exceeds a certain threshold, an  excited cell becomes refractory, and a refractory cell returns to its original 
resting state.  In~\cite{adamatzky_holland_1998} we introduced a bit more exotic cellular automaton, 
where a resting cell is excited if a number of its excited neighbours belongs to some fixed 
interval $[\theta_1, \theta_2]$. The interval $[\theta_1, \theta_2]$ is called an excitation interval. 
For a two-dimensional cellular automaton with eight-cell neighbourhood boundaries of the excitation interval are between 1 and 8: $1 \leq \theta_1 \leq \theta_2 \leq 8$. By tuning $\theta_1$ and $\theta_2$ we control automata dynamics and evoke target and spiral waves, stationary excitation patterns, and mobile localisaitons~\cite{adamatzky_book_2005}. 

How does excitation dynamics change if we allow boundaries of the excitation interval to change during the automaton development? We partially answered the question in \cite{adamatzky_updatableinterval} by making the interval 
$[\theta_1^t(x), \theta_2^t(x)]$ of every cell $x$  dynamically updatable at every step $t$ depending on state of the cell $x$ and numbers of excited and refractory neighbours in the cell $x$'s neighbourhood. We found that  
excitable cellular automata with dynamical excitation intervals exhibit a wide range of space-time dynamics based 
on an interplay between  propagating excitation patterns and excitability of cells modified by the excitation patterns. Such interactions lead to formation of standing domains of excitation, stationary waves and localised excitations. We analysed  morphological and generative diversities of the functions studied and characterised the functions with 
highest values of  the diversities.  

Excitable cellular automata with dynamical intervals of excitation can be considered as discrete phenomenological models, or rather conceptual analogs, of memristive media and excitable chemical medium computers.  

\subsection{Memristive medium}

The memristor --- a passive resistor with memory --- is a device whose resistance changes depending on the 
polarity and magnitude of a voltage applied to the device's terminals and the duration of this voltage's application. 
Its existence was theoretically postulated by Leon Chua in 1971 based on symmetry in integral variations of 
OhmÕs laws~\cite{chua:1971,chua:1976,chua:1980}.    The first experimental prototypes of memristors are 
reported in~\cite{williams:2008,erokhin:2008,yang:2008}.  An importance, and the great pragmatic value, 
of memristors is that one can design logically universal, or functionally complete,  circuits composed entirely of the memristors. Potential unique applications of memristors are in spintronic devices, ultra-dense information storage, 
neuro-morphic circuits~\cite{erokhin_2009}, and programmable electronics~\cite{strukov:2008}, designing binary arithmetical circuits with polymer organic memristors~\cite{erokhin_howard_adamatzky}.

Despite phenomenal number of results in  memristors produced literally every week there are insufficient 
findings on phenomenology of spatially extended non-linear media with hundreds of thousands of locally 
connected memristive elements. Three cellular automaton models of a memristive medium were suggested so far:
\begin{itemize}
\item  Itoh-Chua memristor cellular automata, where cellular automaton lattice is actually 
designed of memristors~\cite{itoh:2009},
\item Adamatzky-Chua model of memrisitive cellular automata based on structurally-dynamic 
cellular automata~\cite{adamatzky_memristive_excitable},
\item  semi-memristive automata~\cite{adamatzky_chua_semimemristive}.
\end{itemize}

Itoh-Chua and Adamatzky-Chua models imitate memristive properties of links, connections 
between cells of automata arrays but not the cells themselves. The semi-memristive automata 
bring memristivity into cells~\cite{adamatzky_chua_semimemristive}: links between cells are 
always 'conductive' but cells themselves can take non-conductive, or refractory, states. 
The semi-memristive automata are excitable cellular automata with retained refractoriness. 

In present paper, an excitability of a cell is calculated as a number of possible neighbourhood 
configurations that excite the resting cell. \emph{Cells with maximal values of excitability are called conductive.} 
We represent conductivity of a cell $x$ via boundaries  of its excitation interval 
$[\theta_1(x), \theta_2(x)]$. We excite the cellular automaton, wait till the perturbation spreads
and boundaries of excitation intervals of cells updated and then select cells with highest values of 
excitability. The configurations of conductive cells form conductive pathways by analogy with the
formation of conductive pathways in disordered networks of organic memristors~\cite{erokhin_2010}.
In automata studied polarity of a direct current applied to a cell $x$ 
is imitated by excited and refractory neighbours of $x$, and current intensity is represented by a ratio 
of excited and refractory neighbours.

\subsection{Ensembles of Belousov-Zhabotinsky vesicles}

Excitable reaction-diffusion computers, especially those based on Belousov-Zhabotinsky reaction, 
employ principles of a collision-based computing~\cite{adamatzky_book_2005}. Wave-fragments collide in a 'free' 
space and change their velocity vectors in the result the collision. When input and output waves
are interpreted as logical variables., the site of the waves' collision can be seen as a logical 
gates. Wave-fragments, similar to dissipative solitons~\cite{bode_2002}, are 
inherently unstable: they either collapse or explode. A way to overcome the problem of wave-fragments' instability
was suggested in \cite{grunert_2011, adamatzky_biosystems_2012, king_2012}: 
a subdivision of the computing substrate into interconnected compartments, so called BZ-vesicles,
and allowing waves to collide only inside the compartments. Each BZ-vesicle has a membrane that is impassable for excitation. A pore, or a channel, between two vesicles is formed when two vesicles come into direct contact.  The pore is small such that when a wave passes through the pore there is insufficient time for the wave to expand or collapse before interacting with other waves entering through adjacent pores, or sites of contact. 

It has been observed in chemical laboratory experiments with BZ-vesicles~\cite{Szymanski_2011} that 
waves of oxidation induced by external stimulation, e.g. with a silver wire, propagating in the initially 
resting BZ medium may cause changes in the excitability of BZ-vesicles (not just refractoriness but excitability in a long run). Excitability of BZ-vesicles can increase after first round of the oxidation wave propagation.  
A cellular automaton model designed in present paper gives a phenomenological snapshot of an ensemble of regularly 
arranged BZ-vesicles (imitated by cells), which change their long-term excitability after being subjected to propagating waves 
of excitation.

The paper is structured as follows. We define an excitable cellular automata with dynamically updated boundaries of excitation interval in Sect.~\ref{definition}. Configurations of conductivity generated by the automata are analysed in 
Sect.~\ref{conductiveconfigurations}. The functions which produce fully conductive configurations are selected in exhaustive search.  Section~\ref{designingwires} demonstrates how to design conductive wire-like pathways by positions seeds of excitation. Potential further developments are outlined in Sect.~\ref{discussion}.

\section{Excitation controlled excitation intervals}
\label{definition}

Let $L$ be a two-dimensional orthogonal array of finite state machines, or cells, $x^t$ and $x^{t+1}$ 
 states of a cell $x$ at time steps $t$ and $t+1$, and  $\sigma^t_+(x)$ a sum of 
excited neighbours in cell $x$'s eight-cell neighbourhood $u(x)=\{ y : |x - y |_{L_\infty} = 1 \}$.  
Cell $x$ updates its state by the following rule, $x^{t+1}=f(u(x)$, where cell-state update 
function $f$ is represented as 
\begin{equation}
x^{t+1}=
\begin{cases}
+, \text{ if } x^t=\circ \text{ and }  \sigma_+^t(x)+ \in [\theta_1^t(x), \theta_2^t(x)] \cr
-, \text{ if } x^t=+ \cr
\circ, \text{ otherwise } 
\end{cases}
\end{equation}
A resting cell is excited if a number of its neighbours belongs to excitation interval $[\theta_1^t(x), \theta_2^t(x)]$,  
where $1 \leq \theta_1^t(x), \theta_2^t(x) \leq 8$.  The boundaries $\theta_1^t(x)$  and $\theta_2^t(x)$ are 
dynamically updated depending on cell $x$'s state and numbers of $x$'s excited $\sigma^t_+(x)$ and 
refractory $\sigma^t_-(x)$ neighbours. A natural way to update boundaries is by increasing or decreasing
their values as follows: 
\begin{equation}
\begin{array}{cc}
\theta_1^{t+1}(x) = \xi(\theta_1^t(x) + \Delta_1 \phi(\sigma^t_+(x) - \sigma^t_-(x))) \\
\theta_2^{t+1}(x) = \xi(\theta_2^t(x) + \Delta_2 \phi(\sigma^t_+(x) - \sigma^t_-(x))) 
\end{array}
\end{equation}
where
\begin{equation}
\begin{array}{cc}
\Delta_1 = 
\begin{cases}
T_1, \text{ if } x=+ \\
T_3, \text{ if } x=- \\
0, \text{ if } x=0
\end{cases}
&
\Delta_2 = 
\begin{cases}
T_2, \text{ if } x=+ \\
T_4, \text{ if } x=- \\
0, \text{ if } x=0
\end{cases}
\end{array}
\end{equation}
and $\phi(a - b)=1$ if $a>b$, 0 if $a=b$ and -1 if $a<b$, and  $\xi(a)=1$ if $a<1$ and 8 if $a>8$.  
Boundaries of excitation interval $[\theta_1^t(x), \theta_2^t(x)]$ are updated independently of each other. 
 Rules of excitation intervals update are determined by values of $T_1, \cdots T_4$. We therefore address the functions as tuples $E(T_1, T_2, T_3, T_4)$ which range from $E(-1,-1,-1,-1)$ to $E(1,1,1,1)$.

Excitability $\mathcal{E}(\theta_1(x), \theta_2(x))$ of a cell $x$  with excitation interval $[\theta_1(x), \theta_2(x)]$ 
is measured as a number of all possible local configurations which have sum of excited cells lying in the excitation interval $[\theta_1(x), \theta_2(x)]$: 
\begin{equation}
\mathcal{E}(\theta_1, \theta_2) = |\{  w \in \{ \circ, +, - \}^8\}: f(w)=+|
\end{equation}
Two highest excitability values are reached by cells with excitation intervals $[1,7]$  and $[1,8]$,
$\mathcal{E}(1,7)=6304$ and $\mathcal{E}(1,8)=6305$. 
We assume a cell $x$ is conductive if $\mathcal{E}(\theta_1(x), \theta_2(x))>6300$.  
Connectivity of the configurations of cells with maximal excitability is considered to 
be analog of conductivity of the whole cellular array. The connectivity is estimated using standard 
bucket fill algorithm. 

\begin{definition}
We call conductivity configuration diameter $D$ fully conductive if there is a path between two sites lying 
at distance $D$ from each other, and over 90\% of sites are reached from any given site.  
\end{definition}

We study formation of conductive pathways in initially non-conductive medium. Therefore, 
in experiments we considered initial conditions $\theta^0_1(x)=2$ and  $\theta^0_2(x)=8$ for any $x$. 
The excitation interval $[2,8]$ gives a cell excitability 5281, which is below a threshold adopted as indicator of
conductivity.   

We excite a disc radius 200 cells with a random configuration of excited states. Let $p$ be a probability for a 
cell $x$ to be assigned excited state at $t=0$ and one of its neighbours, chosen at random, also assigned 
an excited state.  Two neighbouring excited cells is a minimal size of perturbation due to  $\theta^0_1(x)=2$. 
We considered $p=10^{-3}$ and $p=0.1$. Initially cells inside a disc radius 200 are assigned excited states  
with probability $p$. 

In each trial we allowed the cellular automaton to develop for $440$ iterations and then analysed configurations of 
conductivity.  Size of cellular arrays was chosen large enough for a front of perturbation to never reach
the array's boundaries in $440$ steps, so there are no influences of boundary conditions.

\section{Conductive configurations}
\label{conductiveconfigurations}

\begin{figure}[!tbp]
\centering
%\subfigure[Excitation]{\label{21excitation} \includegraphics[scale=0.2]{figs/21_excitation}}
%\subfigure[$\theta_1$]{\label{21theta1} \includegraphics[scale=0.2]{figs/21_theta_1}}
%\subfigure[$\theta_2$]{\label{21theta2} \includegraphics[scale=0.2]{figs/21_theta_2}}
%\subfigure[Conductivity]{\label{21conductivity} \includegraphics[scale=0.24]{figs/21_conductivity}}
\includegraphics[width=\textwidth]{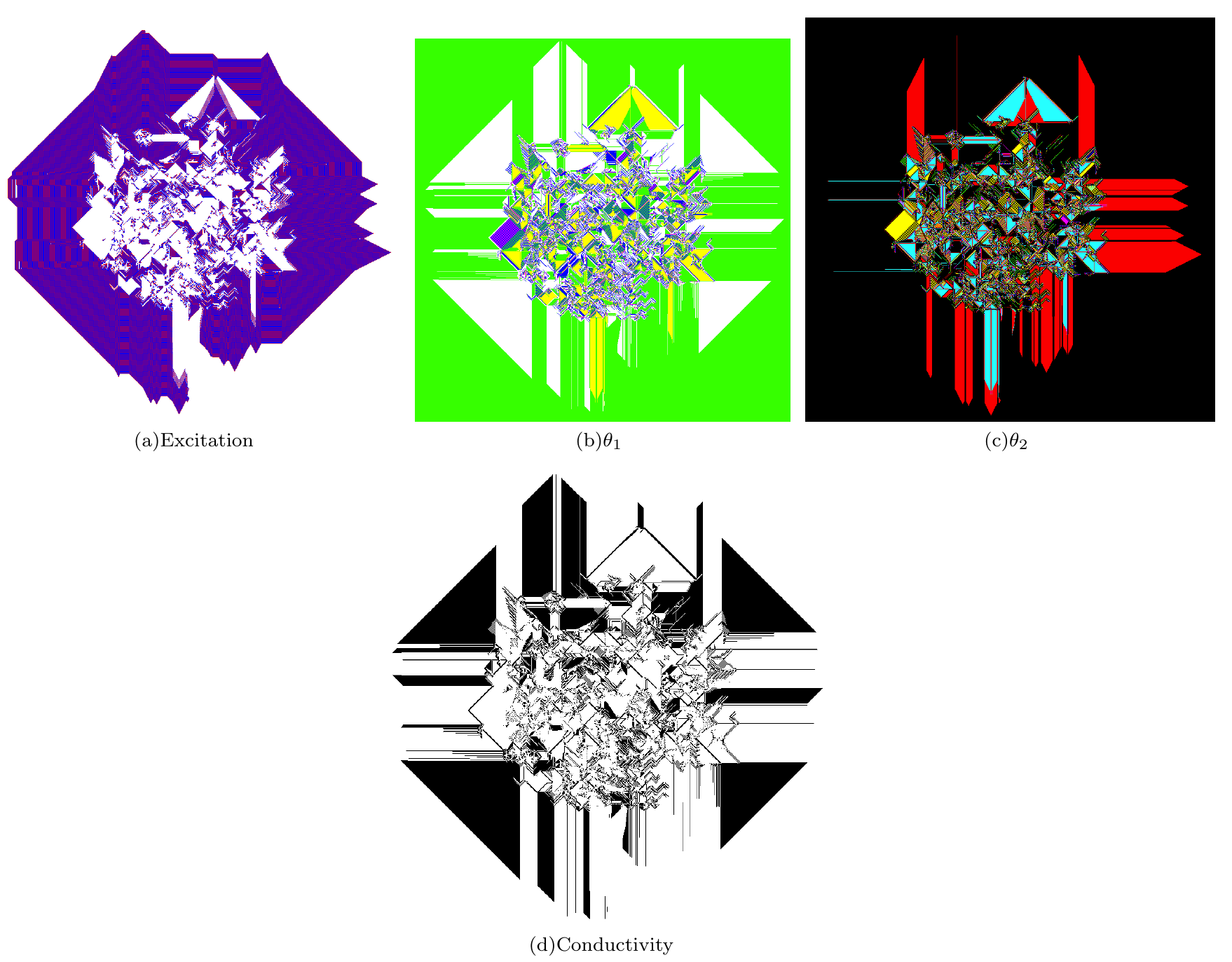}
\caption{(Color online) Snapshots of (a)~excitation pattern, (b)~configuration of lower boundary $\theta_1$ of excitation 
interval, (c)~configuration of upper boundary $\theta_2$ of excitation interval, (d)~configuration of conductivity of
cellular automaton governed by function $E(-1,1,0,-1)$. Array of $880 \times 880$ cells evolved for 440 steps. 
Initially cells inside a disc radius 200 are assigned excited states with probability $10^{-3}$. Cell states  in (a) are represented by colours and grey levels as follows: excited state $+$ is red (c. 76 grey), resting state is white and refractory state $-$ is blue (c. 28 grey). Colour values of  excitation interval boundaries $\theta_1$~(b) and 
$\theta_2$~(c)  are following: 1 is white, 2 is green or   150 grey, 3 is yellow or 226 grey, 4 is blue or 28 grey, 
5 is magenta or 104 grey,  6 is cyan or 178 grey, 7 is red or 76 grey, and 8 is black. Conductive cells in (d) are black and non-conductive are white. 
}
\label{21_confs}
\end{figure}

\begin{figure}[!tbp]
\centering
\includegraphics[width=0.7\textwidth]{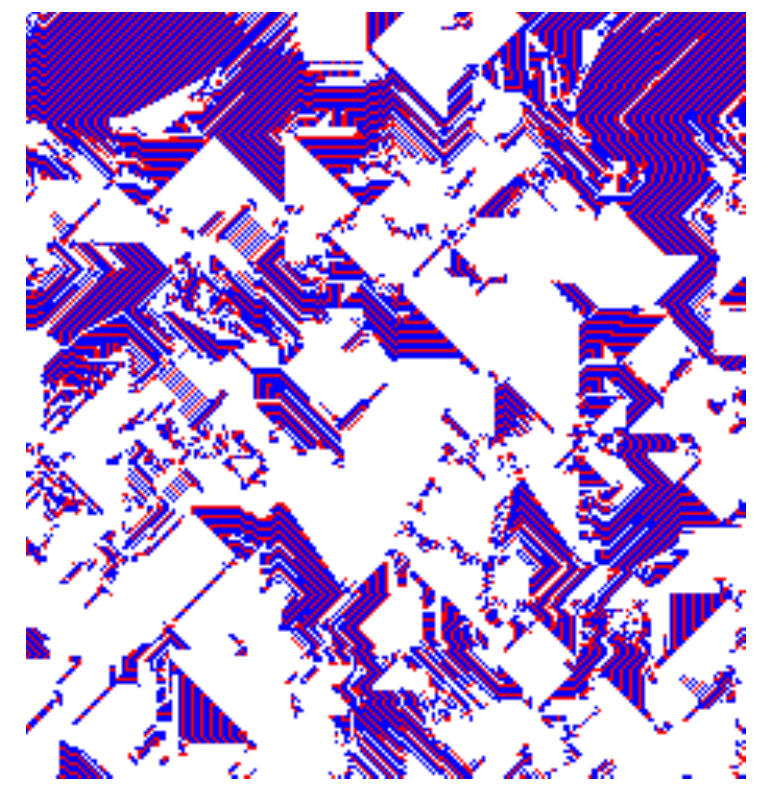}
\caption{(Color online) Enlarged part of excitation pattern from  Fig.~\ref{21_confs}a}
\label{excitationenlarged}
\end{figure}

Sites of initial random excitations generate waves, localizations and other travelling patterns of excitation. 
The patterns either stay localised or merge and propagate outwards the initially perturbed region. The patterns of excitation update boundaries of excitation intervals of cells they excited. An example of cellular automaton, 
function $E(-1,1,0,-1)$, development is shown in Fig.~\ref{21_confs}. Waves of excitation propagate from the perturbation sites (Figs.~\ref{21_confs}a and \ref{excitationenlarged}), leaving somewhat fibre-like trails and extended domains of lower $\theta_1$  (Fig.~\ref{21_confs}b) and upper $\theta_2$  (Fig.~\ref{21_confs}c) boundaries of excitation interval. These are reflected in 
solid domains of conductivity partially linked with wire-like conductive paths (Fig.~\ref{21_confs}d). Excitation waves  originated from different sites of perturbation merge outside the stimulation disc and propagate further as almost connected packet of target waves (Fig.~\ref{21_confs}a). These waves leave triangular solid domains of conductivity behind (Fig.~\ref{21_confs}d).

\subsection{Classes of connectivity}

\begin{figure}[!tbp]
\centering
%\subfigure[$E(1,0,0,-1)$, $\mathcal{N}=(0,4)$]{\label{66exm} \includegraphics[scale=0.2]{figs/(66)100-1}}
%\subfigure[$E(0,-1,1,1)$, $\mathcal{N}=(2,5)$]{\label{35exm} \includegraphics[scale=0.2]{figs/(35)0-111}}
%\subfigure[$E(1,1,0,1)$, $\mathcal{N}=(2,6)$]{\label{77exm} \includegraphics[scale=0.2]{figs/(77)1101}}
%\subfigure[$E(1,1,1,1)$, $\mathcal{N}=(2,7)$]{\label{80exm} \includegraphics[scale=0.2]{figs/(80)1111}}
%\subfigure[$E(0,-1,1,-1)$, $\mathcal{N}=(2,8)$]{\label{33exm} \includegraphics[scale=0.2]{figs/(33)0-11-1}}
%\subfigure[$E(1,-1,0,1)$, $\mathcal{N}=(5,0)$]{\label{59exm} \includegraphics[scale=0.2]{figs/(59)1-101}}
\includegraphics[width=\textwidth]{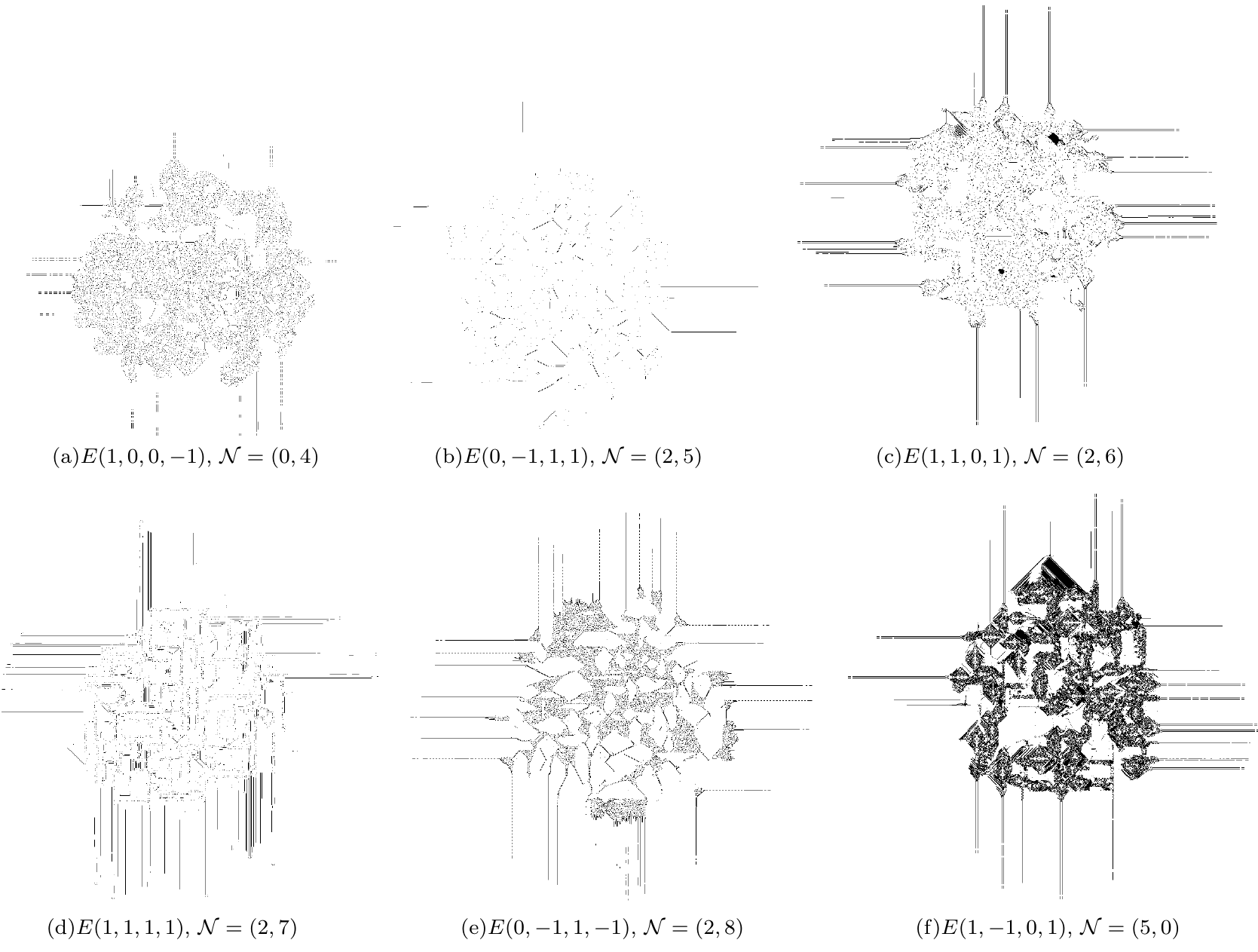}
\caption{Examples of conductivity configurations.}
\label{examplesconfigs1}
\end{figure}

\begin{figure}[!tbp]
\centering
%\subfigure[$E(1,-1,0,0)$, $\mathcal{N}=(7,0)$]{\label{58exm} \includegraphics[scale=0.2]{figs/(58)1-100}}
%\subfigure[$E(1,-1,-1,1)$, $\mathcal{N}=(8,0)$]{\label{56exm} \includegraphics[scale=0.2]{figs/1-1-11(56)exm}}
%\subfigure[$E(1,0,-1,1)$,  $\mathcal{N}=(8,1)$]{\label{65exm} \includegraphics[scale=0.2]{figs/(65)10-11}}
%\subfigure[$E(-1,-1,0,0)$, $\mathcal{N}=(8,3)$]{\label{4exm} \includegraphics[scale=0.2]{figs/(4)-1-100}}
%\subfigure[$E(-1,-1,0,-1)$, $\mathcal{N}=(8,6)$ ]{\label{3exm} \includegraphics[scale=0.2]{figs/(3)-1-10-1}}
\includegraphics[width=\textwidth]{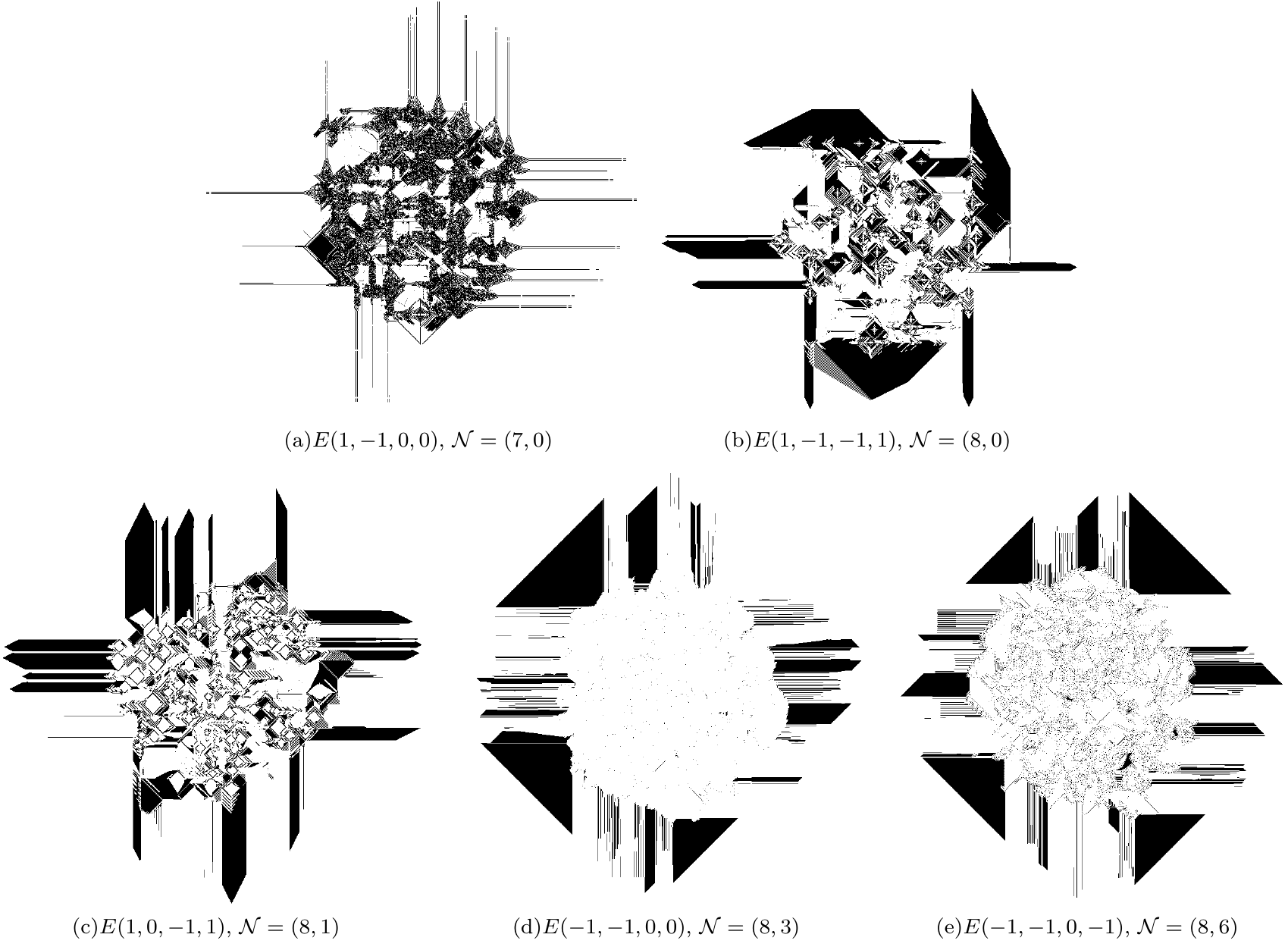}
\caption{Examples of conductivity configurations.}
\label{examplesconfigs2}
\end{figure}

We characterise local connectivity of conductivity configurations using $\nu_{\max}$, 
a number of  conductive neighbours of a conductive cell occurred most frequently in 
the configuration of conductivity, and $\nu_{\min}$, occurred less frequently.  For initial probability of 
excitation $10^{-3}$ we have  twelve classes of local connectivity, the functions are grouped by 
values $\mathcal{N}=(\nu_{\max}, \nu_{\min})$ of their conductivity configurations. Examples of configurations
of conductivity generated by functions from these classes are shown in 
Figs.~\ref{examplesconfigs1} and \ref{examplesconfigs2}.

% $\nu_{\max}$ = \maxD
% $\nu_{\min}$ = \minD

\begin{itemize}
\item $\mathcal{N}=(0,0)$: $E(0 ,-1, 0 ,-1)$, $E(0, -1, 0, 0)$, $E(0, -1, 0, 1)$, $E(0, 0, 0, -1)$,
$E(0, 0, 0, 0)$, $E(0, 0, 0, 1)$, $E(0, 1, 0, -1)$, $E(0, 1, 0, 0)$, $E(0, 1, 0, 1)$.
\item  $\mathcal{N}=(0,4)$: $E(1, 0, 0, -1)$, $E(1, 1, 0, -1)$ (Fig.~\ref{examplesconfigs1}a)
\item 	$\mathcal{N}=(2,5)$: $E(0, -1, 1, 1)$ and $E(1, 1, 0, 0)$  (Fig.~\ref{examplesconfigs1}b)
\item $\mathcal{N}=(2,6)$: $E(1, 1, 0, 1)$  (Fig.~\ref{examplesconfigs1}c)
\item $\mathcal{N}=(2,7)$:	
$E(0, 0, 1, 1)$, 
$E(0, 1, 1, 1)$, 
$E(1, -1, 0, -1)$, 
$E(1, -1, 1, -1)$, 
$E(1, -1, 1, 0)$, 
$E(1, -1, 1, 1)$, 
$E(1, 0, 1, -1)$, 
$E(1, 0, 1, 0)$, 
$E(1, 0, 1, 1)$, 
$E(1, 1, 1, -1)$, 
$E(1, 1, 1, 0)$, 
$E(1, 1, 1, 1)$  (Fig.~\ref{examplesconfigs1}d)   
\item	$\mathcal{N}=(2,8)$:	
$E(0, -1, 1, -1)$, 
$E(0, -1, 1, 0)$, 
$E(0, 0, 1, -1)$, 
$E(0, 0, 1, 0)$, 
$E(0, 1, 1, -1)$, 
$E(0, 1, 1, 0)$  (Fig.~\ref{examplesconfigs1}e)
\item	$\mathcal{N}=(5,0)$:	
$E(1, -1, 0, 1)$,
$E(1, 0, 0, 1)$ (Fig.~\ref{examplesconfigs1}f)
\item $\mathcal{N}=(7,0)$:	
$E(1, -1, 0, 0)$, 
$E(1, 0, 0, 0)$  (Fig.~\ref{examplesconfigs2}a)
\item $\mathcal{N}=(8,0)$:	
$E(-1, -1, -1, -1)$, 
$E(-1, -1, 1, 1)$, 
$E(-1, 0, -1, -1)$, 
$E(-1, 0, 0, -1)$, 
$E(-1, 0, 0, 0)$, 
$E(-1, 1, 0, -1)$, 
$E(-1, 1, 0, 0)$, 
$E(-1, 1, 0, 1)$, 
$E(0, -1, -1, -1)$, 
$E(0, -1, -1, 0)$, 
$E(0, -1, -1, 1)$, 
$E(0, 0, -1, 0)$, 
$E(0, 0, -1, 1)$, 
$E(0, 1, -1, 1)$, 
$E(1, -1, -1, -1)$, 
$E(1, -1, -1, 0)$, 
$E(1, -1, -1, 1)$, 
$E(1, 0, -1, 0)$, 
$E(1, 1, -1, 1)$  (Fig.~\ref{examplesconfigs2}b)
\item $\mathcal{N}=(8,1)$:
$E(-1, -1, -1, 0)$, 
$E(-1, -1, -1, 1)$, 
$E(-1, 0, -1, 0)$, 
$E(-1, 0, -1, 1)$, 
$E(-1, 1, -1, 0)$, 
$E(-1, 1, -1, 1)$, 
$E(1, 0, -1, 1)$  (Fig.~\ref{examplesconfigs2}c)
\item $\mathcal{N}=(8,3)$:
$E(-1, -1, 0, 0)$, 
$E(-1, -1, 0, 1)$, 
$E(-1, -1, 1, 0)$, 
$E(-1, 0, 1, 1)$, 
$E(-1, 1, -1, -1)$, 
$E(-1, 1, 1, 1)$, 
$E(0, 0, -1, -1)$, 
$E(0, 1, -1, -1)$, 
$E(0, 1, -1, 0)$, 
$E(1, 0, -1, -1)$, 
$E(1, 1, -1, -1)$, 
$E(1, 1, -1, 0)$  (Fig.~\ref{examplesconfigs2}d)
\item $\mathcal{N}=(8,6)$:
$E(-1, -1, 0, -1)$, 
$E(-1, -1, 1, -1)$, 
$E(-1, 0, 0, 1)$, 
$E(-1, 0, 1, -1)$, 
$E(-1, 0, 1, 0)$, 
$E(-1, 1, 1, -1)$, 
$E(-1, 1, 1, 0)$  (Fig.~\ref{examplesconfigs2}e)
\end{itemize}

Class $\mathcal{N}=(8,0)$ is the largest class, it has 19 functions. Configurations of conductivity generated by the functions from $\mathcal{N}=(8,0)$ (Fig.~\ref{examplesconfigs2}b) are characterised by large solid conductive domains formed either by initial 
source of excitation, see e.g. rectangular embedded shapes in  Fig.~\ref{examplesconfigs2}b, or by merging fronts 
of excitation (solid polygonal domains at the periphery). 

The next largest classes are $\mathcal{N}=(2,7)$ and 
$\mathcal{N}=(8,3)$ (Fig.~\ref{examplesconfigs2}d).  Each of them includes 12 functions. Conductivity configurations generated
by functions of class $\mathcal{N}=(2,7)$  consist mainly of isolated line segments of cells in conductive states and 
pairs or singletons.  Configurations generated by functions of $\mathcal{N}=(8,3)$ are comprised of solid domains of conductive states and sparsely scattered singletons (Fig.~\ref{examplesconfigs1}d). 

Classes $\mathcal{N}=(2,5)$ (Fig.~\ref{examplesconfigs1}b), $\mathcal{N}=(2,6)$ (Fig.~\ref{examplesconfigs1}c) and $\mathcal{N}=(2,7)$ (Fig.~\ref{examplesconfigs1}d) pose particular interest because majority of conductive cells have two conductive neighbours each and therefore chances of conductive 'wires' to be formed could be high. 

\begin{figure}[!tbp]
%\subfigure[]{\label{fragment35} \includegraphics[scale=0.1]{figs/fragment35}}
%\subfigure[]{\label{fragment77} \includegraphics[scale=0.1]{figs/fragment77}}\\
%\subfigure[]{\label{fragment33} \includegraphics[scale=0.3]{figs/fragment33}}
\includegraphics[width=0.5\textwidth]{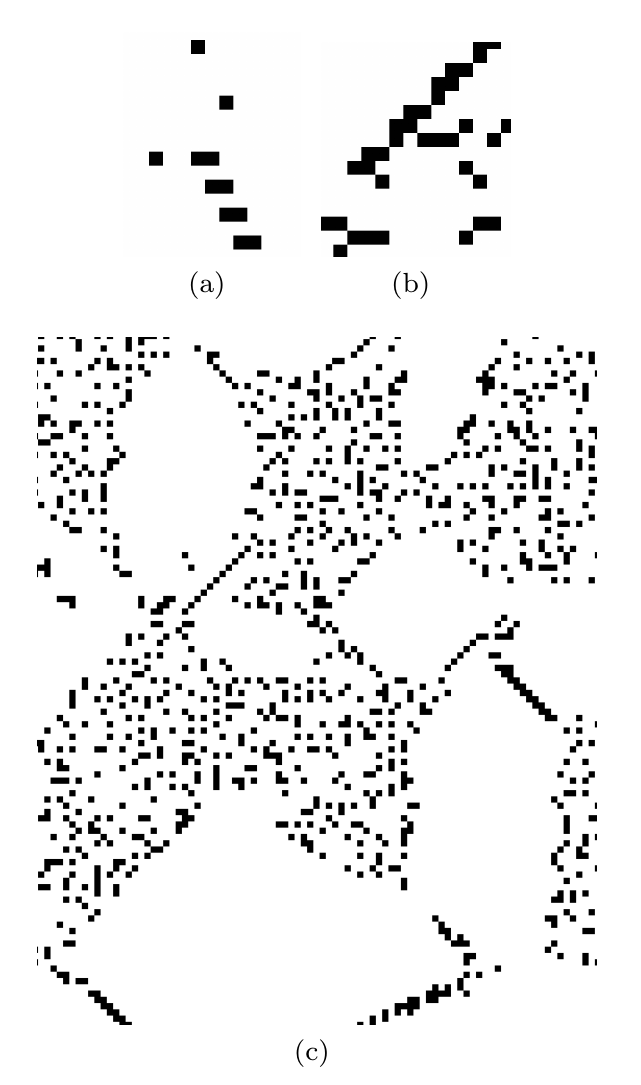}
\caption{Fragments of configurations generated by (a)~function $E(0,-1,1,1)$ from $\mathcal{N}=(2,5)$ , 
(b)~function $E(1,1,0,1)$ from $\mathcal{N}=(2,6)$, (c)~function $E(0,1,1,0)$ from $\mathcal{N}=(2,8)$. }
\label{zoomed}
\end{figure}

Configurations generated by functions from $\mathcal{N}=(2,5)$  (Fig.~\ref{examplesconfigs1}b) predominantly consist of singletons and pairs of cells  in conductive states; the pairs are aligned in staircase like fashion but there is no immediate connection 
between the pairs (Fig.~\ref{zoomed}a).  There are zig-zag like 'wires' of conductive states  (Fig.~\ref{zoomed}b) 
in configurations  (Fig.~\ref{examplesconfigs1}c)  generated by the only function  $E(1,1,0,1)$ in class $\mathcal{N}=(2,6)$. However they are present as isolated fragments. Configurations (Fig.~\ref{examplesconfigs1}e) generated by function $E(0, 1, 1, 0)$ of class 
$\mathcal{N}=(2,8)$ consist of clusters of singletons --- single conductive cells surrounded by non-conductive neighbours ---
and pairs of cells in conductive states. The clusters are connected to each other with diagonals of singletons 
 (Fig.~\ref{zoomed}c). Neither of the functions with $\nu_{max}=2$ generate fully conductive configurations but few functions from classes $\mathcal{N}=(7,0)$ and $\mathcal{N}=(8,0)$ do. We study these functions below.

\subsection{Functions generating conductive configurations}

\begin{figure}[!tbp]
\centering
%\subfigure[$E(1,-1,0,1)$]{\label{59exmColored} \includegraphics[scale=0.4]{figs/(59)1-101COLOURED}}
%\subfigure[$E(-1,1,0,0)$]{\label{22exmLow} \includegraphics[scale=0.4]{figs/(22)-1100LowProb}}
%\subfigure[$E(1,0,0,1)$]{\label{68exmLow} \includegraphics[scale=0.4]{figs/(68)1001LowProb}}
\includegraphics[width=\textwidth]{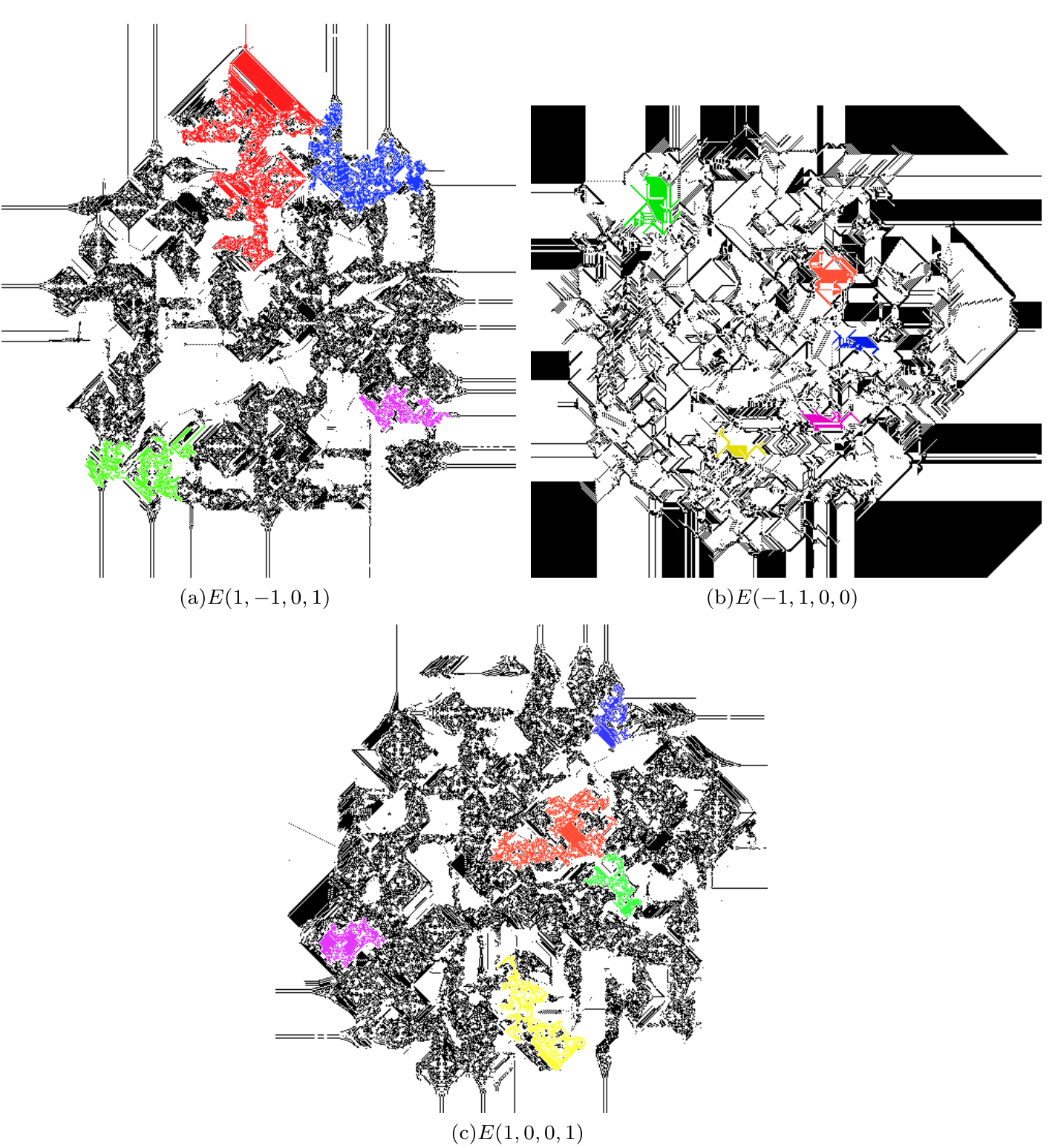}
\caption{(Color online)  Examples of configurations with partially connectivity clusters coloured. The clusters are bucket flooded with red, blue, green, magenta and yellow colours (seen as shades of grey in black and white reproduction).}
\label{conductancecoloring6}
\end{figure}

Most functions generate only partially conductive configurations of conductivity. Examples of connected clusters are shown in Fig.~\ref{conductancecoloring6}.

\begin{finding}
For low probability of initial excitation, $p=10^{-3}$, functions $E(1,0,0,0)$ and $E(1,-1,0,0)$ generate fully conductive 
configurations. For high probability of initial excitations, $p=0.1$, fully conductive configurations are generated by functions $E(-1,0,0,0)$ and $E(-1,1,0,0)$.
\end{finding}

\begin{figure}[!tbp]
\centering
%\subfigure[$E(1,0,0,0)$]{\label{58exmColored} \includegraphics[scale=0.4]{figs/(58)1-100COLOURED}}
%\subfigure[$E(1,-1,0,0)$]{\label{67exmColored} \includegraphics[scale=0.4]{figs/67Coloured}}
%\subfigure[$E(-1,0,0,0)$]{\label{13exmHigh} \includegraphics[scale=0.4]{figs/(13)-1000High}}
%\subfigure[$E(-1,1,0,0)$]{\label{22exmHigh} \includegraphics[scale=0.4]{figs/(22)-1100High}}
\includegraphics[width=\textwidth]{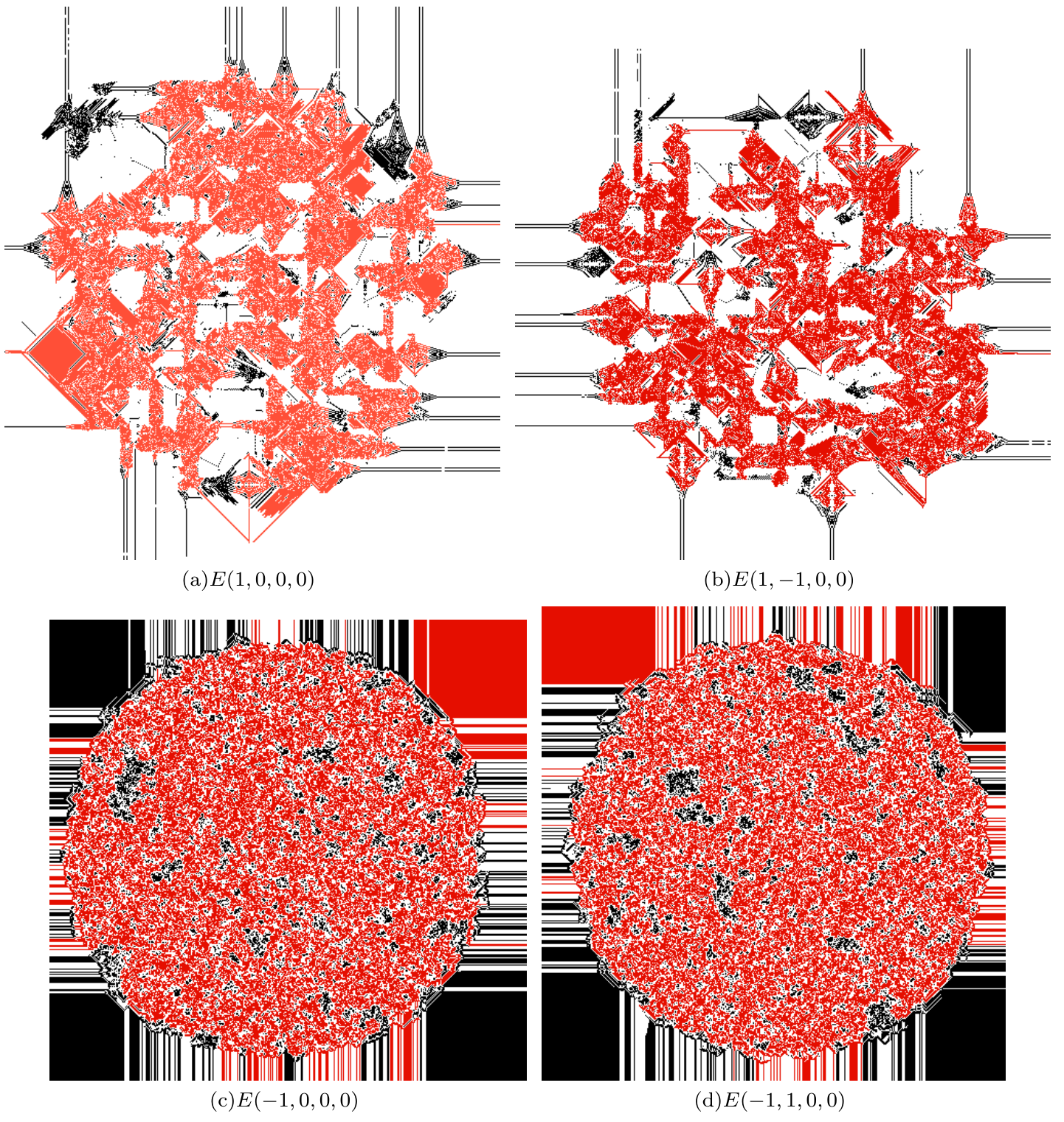}
\caption{(Color online)  Examples of conductivity configurations generated by functions (a)~$E(1,0,0,0)$,
(b)~$E(1,-1,0,0)$, (c)~$E(-1,0,0,0)$ and (d)~$E(-1,1,0,0)$. Non-conductive cells are white. Instances of 
connected clusters of conductive cells are bucket flooded with red (c. 76 grey).}
\label{conductancecoloring}
\end{figure}

\begin{figure}[!tbp]
%\begin{tabular}{c||c|c|c|c|}
% $p$ &   $T_1$ & $T_2$ 	& $\sigma_+ > \sigma_-$ & $\sigma_+ < \sigma_-$ \\  \hline
 % $10^{-3}$ & 1 & 0 & \includegraphics[scale=0.3]{figs/LowerUp} &  \includegraphics[scale=0.3]{figs/LowerDown} \\
  %  & -1 & 1 & \includegraphics[scale=0.3]{figs/IntervalShrinks} &  \includegraphics[scale=0.3]{figs/IntervalExpands} \\ %\hline
  %   $10^{-1}$ & -1 & 0 & \includegraphics[scale=0.3]{figs/LowerDown} &  \includegraphics[scale=0.3]{figs/LowerUp} \\
 %       & -1 & 1 & \includegraphics[scale=0.3]{figs/IntervalExpands} &  \includegraphics[scale=0.3]{figs/%IntervalShrinks} \\
%\end{tabular}
\includegraphics[width=\textwidth]{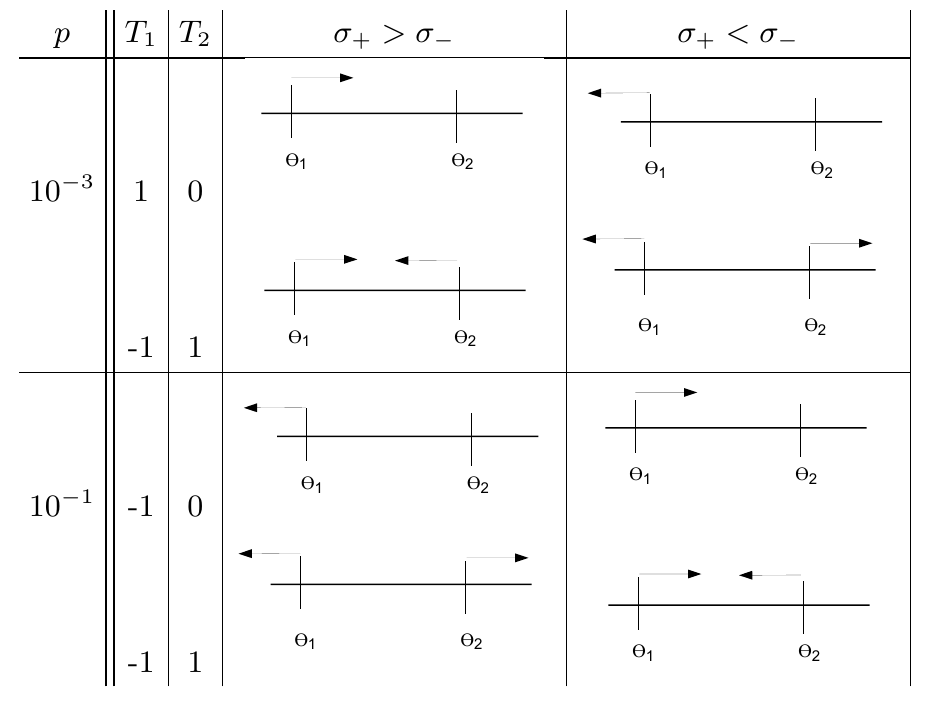}
\caption{Mechanics of excitation interval update in functions  $E(1,0,0,0)$, $E(1,-1,0,0)$, $E(-1,0,0,0)$ and $E(-1,1,0,0)$. 
$\sigma_+$ is a number of excited neighbours in a cell's neighbourhood and $\sigma_-$ is a number of refractory
neighbours.}
\label{explanations}
\end{figure}

For low probability $p=10^{-3}$ of initial excitation only functions  $E(1,-1,0,0)$ (Figs.~\ref{conductancecoloring}a) 
and $E(1,0,0,0)$ %67
generate fully conductive configurations. 
For initial configurations with high probability of excited cells ($p=0.1$) only two functions generate conductive 
configurations: 
$E(-1,0,0,0)$ %13 
(Fig.~\ref{conductancecoloring}c)
and 
$E(-1,1,0,0)$ %22
(Fig.~\ref{conductancecoloring}d).
In all four functions selected values $T_3$ and $T_4$ are nil. The values correspond to update of excitation intervals 
of refractory cells. This is in line with commonly accepted assumption that in an excitable medium elements in refractory state are insensitive to states of their neighbours. Mechanics of excitation interval updates of excited cells
is illustrated in Fig.~\ref{explanations}. 

Function $E(1,0,0,0)$ increases lower boundary $\theta_1(x)$ of excitability interval of a cell $x$ when a  number of 
excited neighbours of the cell exceeds  a number of refractory neighbours. If a cell has less excited neighbours than
refractory ones then lower boundary of the excitability interval decreases.  Function  $E(1,-1,0,0)$ increases lower
boundary $\theta_1$ and decreases upper boundary $\theta_2$ of excitability interval of a cell when excited neighbours of the cells are in majority. If refractory cells dominate in a cell's neighbourhood then $\theta_1$ decreases and 
$\theta_2$ increases (Fig.~\ref{explanations}). Both functions decrease excitability of a cell when the cell's neighbourhood is 'over-excited' (prevalence of excited neighbours) and increase the cell's excitability when the cell's 
neighbourhood 'over-refractory' (prevalence of refractory neighbours). Thus they stabilise the excitation dynamics. 

Function $E(-1,0,0,0)$ decreases $\theta_1$ when a cell has majority of excited neighbours and increases $\theta_1$ if refractory neighbours dominate. Function $E(-1,1,0,0)$ expands excitation interval ($\theta_1$ decreases and $\theta_2$ 
increases) when there are more excited than refractory neighbours; and contracts the excitation interval 
($\theta_1$ increases and $\theta_2$ decreases) when refractory neighbours prevail (Fig.~\ref{explanations}). The functions increase excitability of cells with  'over-excited' neighbourhoods and decrease excitability of cells with 'over-refractory' neighbourhoods.  Thus they destabilise the excitation dynamics.

\begin{finding}
For low probabilities of initial perturbations functions stabilising excitation dynamics generate conductive configurations while for high probabilities of initial perturbations functions de-stabilising excitation dynamics generate conductive configurations. 
\end{finding}

Let $E(e_1, e_2, e_3, e_4)$ be a function generating conductive configurations for low probability of excitation in initial 
configurations and $E'(e'_1, e'_2, e'_3, e'_4)$ the function generating conductive configurations for high probability of 
excitation in initial configuration. Then 
$$
E(e_1, e_2, e_3, e_4) = E'(-1\cdot e'_1, -1\cdot e'_2, -1\cdot e'_3, -1\cdot e'_4)
$$

\section{Growing conductive pathways with functions $E(1-100)$ and $E(1000)$}
\label{designingwires}

%\begin{figure}
%\subfigure[$t=0$]{\input{figs/1000Seed/0E.tex}}
%\subfigure[$t=1$]{\input{figs/1000Seed/1E.tex}}
%\subfigure[$t=2$]{\input{figs/1000Seed/2E.tex}}
%\subfigure[$t=3$]{\input{figs/1000Seed/3E.tex}}
%\subfigure[$t=4$]{\input{figs/1000Seed/4E.tex}}
%\subfigure[$t=5$]{\input{figs/1000Seed/5E.tex}}
%\subfigure[$t=6$]{\input{figs/1000Seed/6E.tex}}
%\subfigure[$t=7$]{\input{figs/1000Seed/7E.tex}}
%\subfigure[$t=8$]{\input{figs/1000Seed/8E.tex}}
%\subfigure[$t=9$]{\input{figs/1000Seed/9E.tex}}
%\subfigure[$t=10$]{\input{figs/1000Seed/10E.tex}}
%\subfigure[$t=11$]{\input{figs/1000Seed/11E.tex}}
%\subfigure[$t=12$]{\input{figs/1000Seed/12E.tex}}
%\subfigure[$t=13$]{\input{figs/1000Seed/13E.tex}}
%\subfigure[$t=14$]{\input{figs/1000Seed/14E.tex}}
%\subfigure[$t=15$]{\input{figs/1000Seed/15E.tex}}
%\subfigure[$t=16$]{\input{figs/1000Seed/16E.tex}}
%\subfigure[$t=17$]{\input{figs/1000Seed/17E.tex}}
%\caption{Snapshots of a growing pattern of excitation developed from a seed of two neighbouring cells
%in excited state. Excited cells are shown by black discs, refractory cells by circles and resting cells by 
%gray dots.}
%\label{excitedseedEvolution}
%\end{figure}

\begin{figure}[!tbp]
\centering
\includegraphics[width=\textwidth]{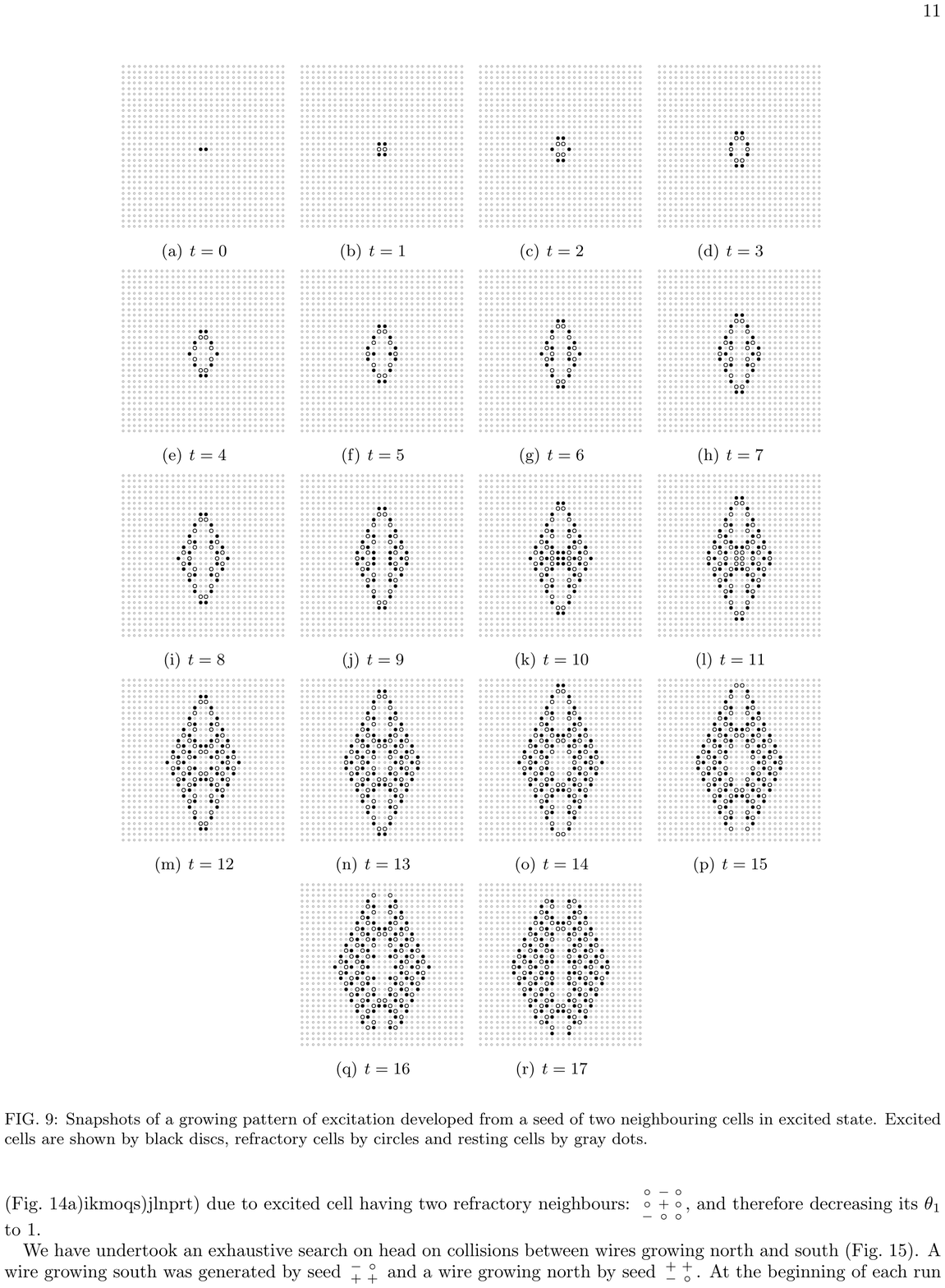}
\caption{Snapshots of a growing pattern of excitation developed from a seed of two neighbouring cells
in excited state. Excited cells are shown by black discs, refractory cells by circles and resting cells by 
gray dots.}
\label{excitedseedEvolution}
\end{figure}

%\begin{figure}
%\subfigure[$t=3$]{\input{figs/1000Seed/3.tex}}
%\subfigure[$t=4$]{\input{figs/1000Seed/4.tex}}
%\subfigure[$t=5$]{\input{figs/1000Seed/5.tex}}
%\subfigure[$t=6$]{\input{figs/1000Seed/6.tex}}
%\subfigure[$t=7$]{\input{figs/1000Seed/7.tex}}
%\subfigure[$t=8$]{\input{figs/1000Seed/8.tex}}
%\subfigure[$t=9$]{\input{figs/1000Seed/9.tex}}
%\subfigure[$t=10$]{\input{figs/1000Seed/10.tex}}
%\subfigure[$t=11$]{\input{figs/1000Seed/11.tex}}
%\subfigure[$t=12$]{\input{figs/1000Seed/12.tex}}
%\subfigure[$t=13$]{\input{figs/1000Seed/13.tex}}
%\subfigure[$t=14$]{\input{figs/1000Seed/14.tex}}
%\subfigure[$t=15$]{\input{figs/1000Seed/15.tex}}
%\subfigure[$t=16$]{\input{figs/1000Seed/16.tex}}
%\subfigure[$t=17$]{\input{figs/1000Seed/17.tex}}
%\caption{Snapshot of configuration of conductive cells developed from seed $(++)$.}
%\label{conductivityEvolution}
%\end{figure}

\begin{figure}[!tbp]
\includegraphics[width=\textwidth]{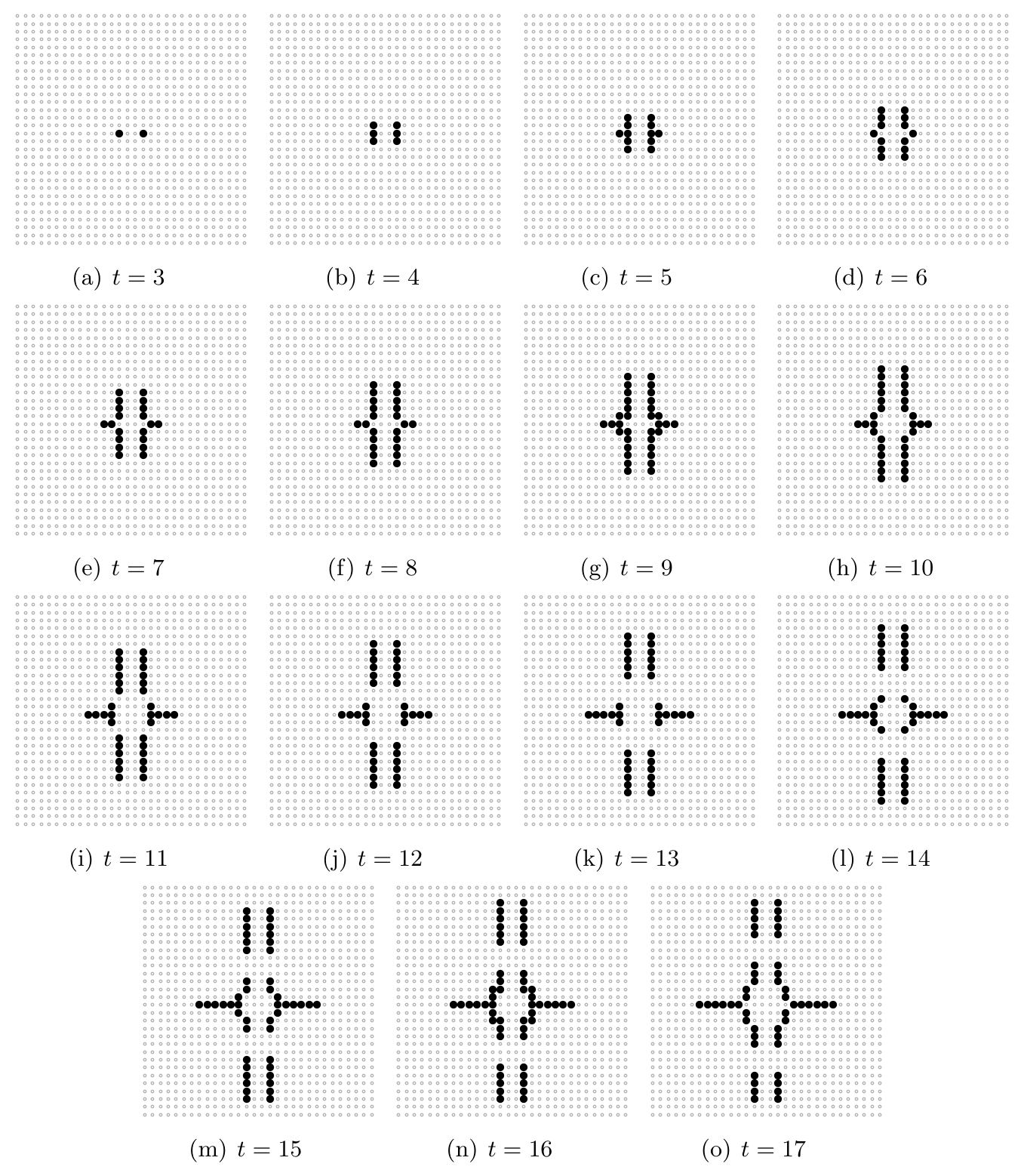}
\caption{Snapshot of configuration of conductive cells developed from seed ++.}
\label{conductivityEvolution}
\end{figure}

Let us consider how to route conductive pathways using minimal resources. The pathways can be initiated by 
 excitating the medium. A minimal seed is a pair of cells in excited state 
 $\begin{smallmatrix} + & + \end{smallmatrix}$ or 
$\begin{smallmatrix} + \\ + \end{smallmatrix}$.  The seed generates a growing pattern of excitation 
(Fig.~\ref{excitedseedEvolution}). The excitation pattern starts as a target wave (Fig.~\ref{excitedseedEvolution}c--f). 
At sixths step of development excitation ruptures inside the wave-front on its eastmost and westmost sites
(Fig.~\ref{excitedseedEvolution}g), and propagates along the wave-front boundary inside the target wave 
(Fig.~\ref{excitedseedEvolution}h--j). At tenth step of the seed's development the excitation reaches the original seed's
position  (Fig.~\ref{excitedseedEvolution}k). Thus the propagating pattern becomes 'filled' with persistent excitation activity.

The associated development of configurations of 
cells in conductive state is shown in Fig.~\ref{conductivityEvolution}. The seed initiates two types of 
growing pathways: single chains and double chains of conductive states. For example, a configuration 
evoked by seed $\begin{smallmatrix} + & + \end{smallmatrix}$  (Fig.~\ref{excitedseedEvolution}a) exhibits two single chains, 
growing west and east, and two double chains, growing north and south  Fig.~\ref{conductivityEvolution}.
The chains growing north and south increase their lengths by one one cell each, speed $1$. 
The chains growing west and east increase by one cell every other step of development, speed $1/2$.

\begin{figure}[!tbp]
\includegraphics[width=0.5\textwidth]{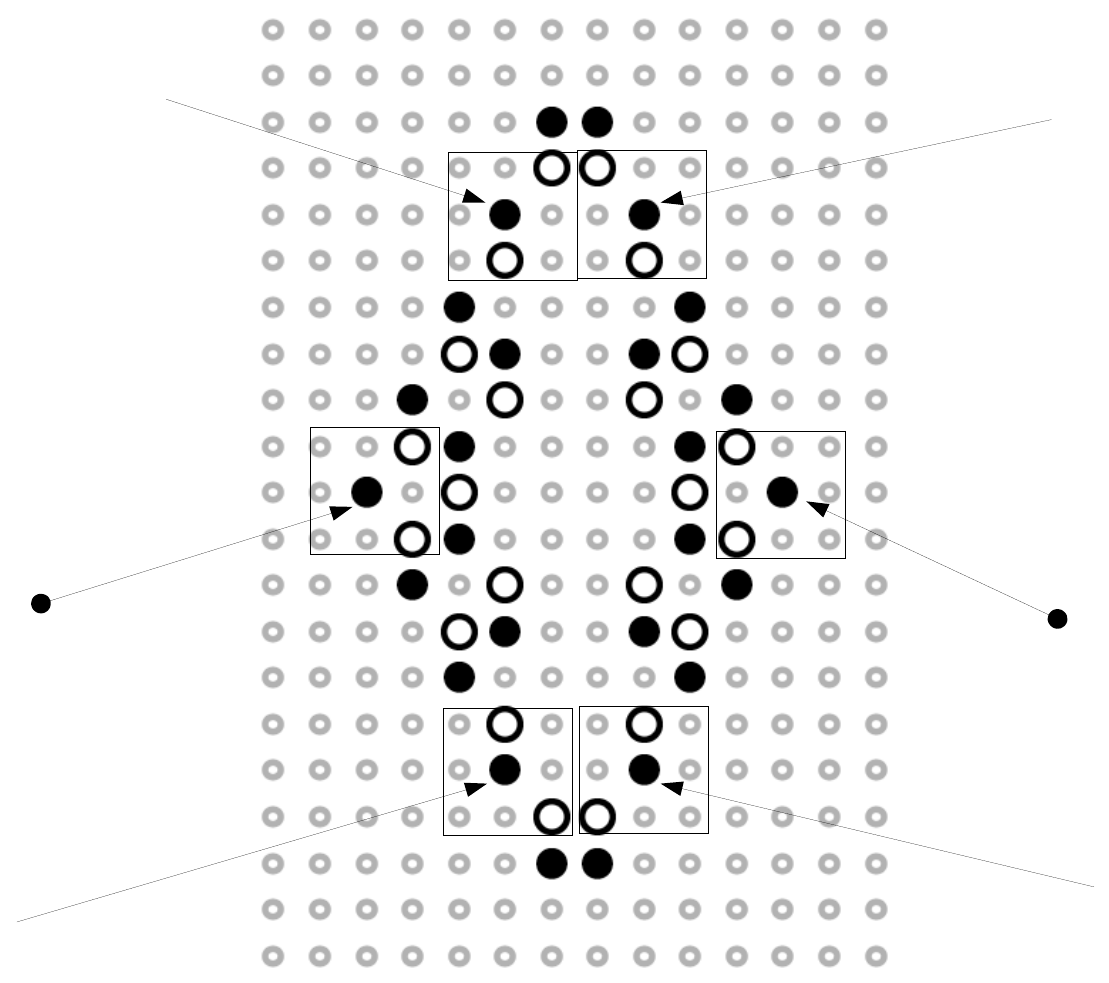}
\caption{Configuration of excited ($\bullet$) and refractory cells ($\circ$) with highlighted cells responsible for formation of conductive pathways. Cells where double wire is formed are marked with arrows and single wire with arrows with disc tops. This configuration develops at sevenths step of the seed's evolution, Fig.~\ref{excitedseedEvolution}h.}
\label{SchemeExplanation}
\end{figure}

The mechanics of formation of conductive pathways is illustrated in Fig.~\ref{SchemeExplanation}. 
Excited cells 
having configurations 
$
\begin{smallmatrix}
\circ & \circ & - \\
\circ & + & \circ \\
\circ & - & \circ 
\end{smallmatrix}
$
and
$
\begin{smallmatrix}
- & \circ & \circ \\
\circ & + & \circ \\
\circ & - & \circ 
\end{smallmatrix}
$ 
(north and south proximities of the excitation pattern), 
and 
configurations
$
\begin{smallmatrix}
\circ & \circ & - \\
\circ & + & \circ \\
\circ & \circ & -
\end{smallmatrix}
$ 
and
$
\begin{smallmatrix}
- & \circ & \circ \\
\circ & + & \circ \\
- & \circ & \circ
\end{smallmatrix}
$ 
(west and east proximities of the excitation patterns)
decrease their values of $\theta_1(x)$ from 2 to 1, thus increasing 
the cells' excitability.

%\begin{figure}
%\subfigure[$t=0$]{\input{figs/EastSouthGrowth/0E.tex}}
%\subfigure[$t=0$]{\input{figs/EastSouthGrowth/0.tex}}
%\subfigure[$t=1$]{\input{figs/EastSouthGrowth/1E.tex}}
%\subfigure[$t=1$]{\input{figs/EastSouthGrowth/1.tex}}
%\subfigure[$t=2$]{\input{figs/EastSouthGrowth/2E.tex}}
%\subfigure[$t=2$]{\input{figs/EastSouthGrowth/2.tex}}
%\subfigure[$t=3$]{\input{figs/EastSouthGrowth/3E.tex}}
%\subfigure[$t=3$]{\input{figs/EastSouthGrowth/3.tex}}
%\subfigure[$t=4$]{\input{figs/EastSouthGrowth/4E.tex}}
%\subfigure[$t=4$]{\input{figs/EastSouthGrowth/4.tex}}
%\subfigure[$t=5$]{\input{figs/EastSouthGrowth/5E.tex}}
%\subfigure[$t=5$]{\input{figs/EastSouthGrowth/5.tex}}
%\subfigure[$t=6$]{\input{figs/EastSouthGrowth/6E.tex}}
%\subfigure[$t=6$]{\input{figs/EastSouthGrowth/6.tex}}
%\subfigure[$t=7$]{\input{figs/EastSouthGrowth/7E.tex}}
%\subfigure[$t=7$]{\input{figs/EastSouthGrowth/7.tex}}
%\caption{XXXXXX see $\begin{array}{cc}  + & \circ \\ + & + \end{array}$ XXX}
%\caption{XXXXXXX}
%\label{EastSouthGrowth}
%\end{figure}

\begin{figure}[!tbp]
\includegraphics[width=\textwidth]{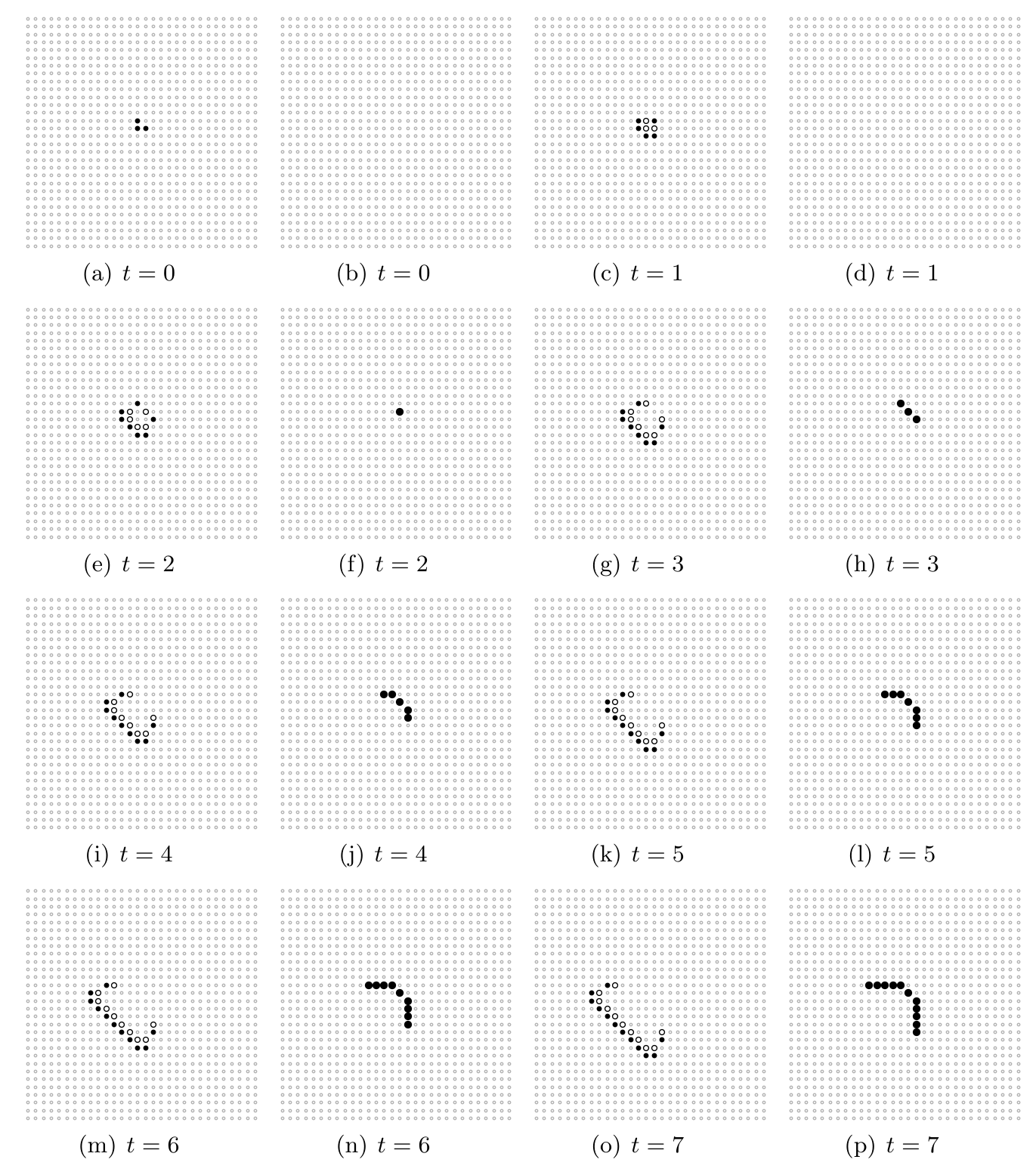}
\caption{Production of two wires growing east and south. For each time step $t$ we show excitation (left pattern) and coductivity (right pattern) configurations. In the excitation configurations excited cells are solid black discs, refractory cells are circles. In the conductivity configurations conductive cells are black discs and non-conductive are grey dots. }
\label{EastSouthGrowth}
\end{figure}

\begin{finding}
By positioning an additional cell in excited or refractory state nearby the original seed of two excited cells it is possible 
to generate a wide range of single thread wires growing into pre-determined directions. 
\end{finding}

Let us consider several examples.

 If we excite a cell above east site of the seed 
$\begin{smallmatrix}   \circ & \circ \\ + & + \end{smallmatrix}$ the modified seed
$\begin{smallmatrix}   + & \circ \\ + & + \end{smallmatrix}$ will develop into a growing 
excitation pattern which generates two chains of cells in conductive states. One chain grows 
east and another chain grows south. Space-time configurations of excited and refractory cells
and conductive cells are shown in Fig.~\ref{EastSouthGrowth}. A seed of three excited cells 
(Fig.~\ref{EastSouthGrowth}a) develops 
into an excitation wave fragment propagating south-east (Fig.~\ref{EastSouthGrowth}egi). The wave-fragment 
expands east and south (Fig.~\ref{EastSouthGrowth}kmo) and resembles an oxidation wave-fragment in 
sub-excitable Belousov-Zhabotinsky medium~\cite{adamatzky_book_2005}. Conductive wires are produced at the sites of excited cells, 
which have neighbourhood configurations 
$
\begin{smallmatrix}
\circ & \circ & \circ \\
\circ & + & - \\
-      & \circ & \circ  
\end{smallmatrix}
$
(north corner of the expanding wave fragment)
and 
$
\begin{smallmatrix}
\circ & - & \circ \\
\circ & + & \circ \\
-      & \circ & \circ  
\end{smallmatrix}
$
(south corner) (Fig.~\ref{EastSouthGrowth}kmo).

%\begin{figure}
%\subfigure[$t=0$]{\input{figs/ThreeGrowth/0E.tex}}
%\subfigure[$t=0$]{\input{figs/ThreeGrowth/0.tex}}
%\subfigure[$t=1$]{\input{figs/ThreeGrowth/1E.tex}}
%\subfigure[$t=1$]{\input{figs/ThreeGrowth/1.tex}}
%\subfigure[$t=2$]{\input{figs/ThreeGrowth/2E.tex}}
%\subfigure[$t=2$]{\input{figs/ThreeGrowth/2.tex}}
%\subfigure[$t=3$]{\input{figs/ThreeGrowth/3E.tex}}
%\subfigure[$t=3$]{\input{figs/ThreeGrowth/3.tex}}
%\subfigure[$t=4$]{\input{figs/ThreeGrowth/4E.tex}}
%\subfigure[$t=4$]{\input{figs/ThreeGrowth/4.tex}}
%\subfigure[$t=5$]{\input{figs/ThreeGrowth/5E.tex}}
%\subfigure[$t=5$]{\input{figs/ThreeGrowth/5.tex}}
%\subfigure[$t=6$]{\input{figs/ThreeGrowth/6E.tex}}
%\subfigure[$t=6$]{\input{figs/ThreeGrowth/6.tex}}
%\subfigure[$t=7$]{\input{figs/ThreeGrowth/7E.tex}}
%\subfigure[$t=7$]{\input{figs/ThreeGrowth/7.tex}}
%\subfigure[$t=8$]{\input{figs/ThreeGrowth/8E.tex}}
%\subfigure[$t=8$]{\input{figs/ThreeGrowth/8.tex}}
%\subfigure[$t=9$]{\input{figs/ThreeGrowth/9E.tex}}
%\subfigure[$t=9$]{\input{figs/ThreeGrowth/9.tex}}
%\caption{XXXXXX see $\begin{array}{ccc}  \circ & \circ & + \\ + & + & \circ \end{array}$ XXX}
%\label{ThreeGrowth}
%\end{figure}

\begin{figure}[!tbp]
\includegraphics[width=\textwidth]{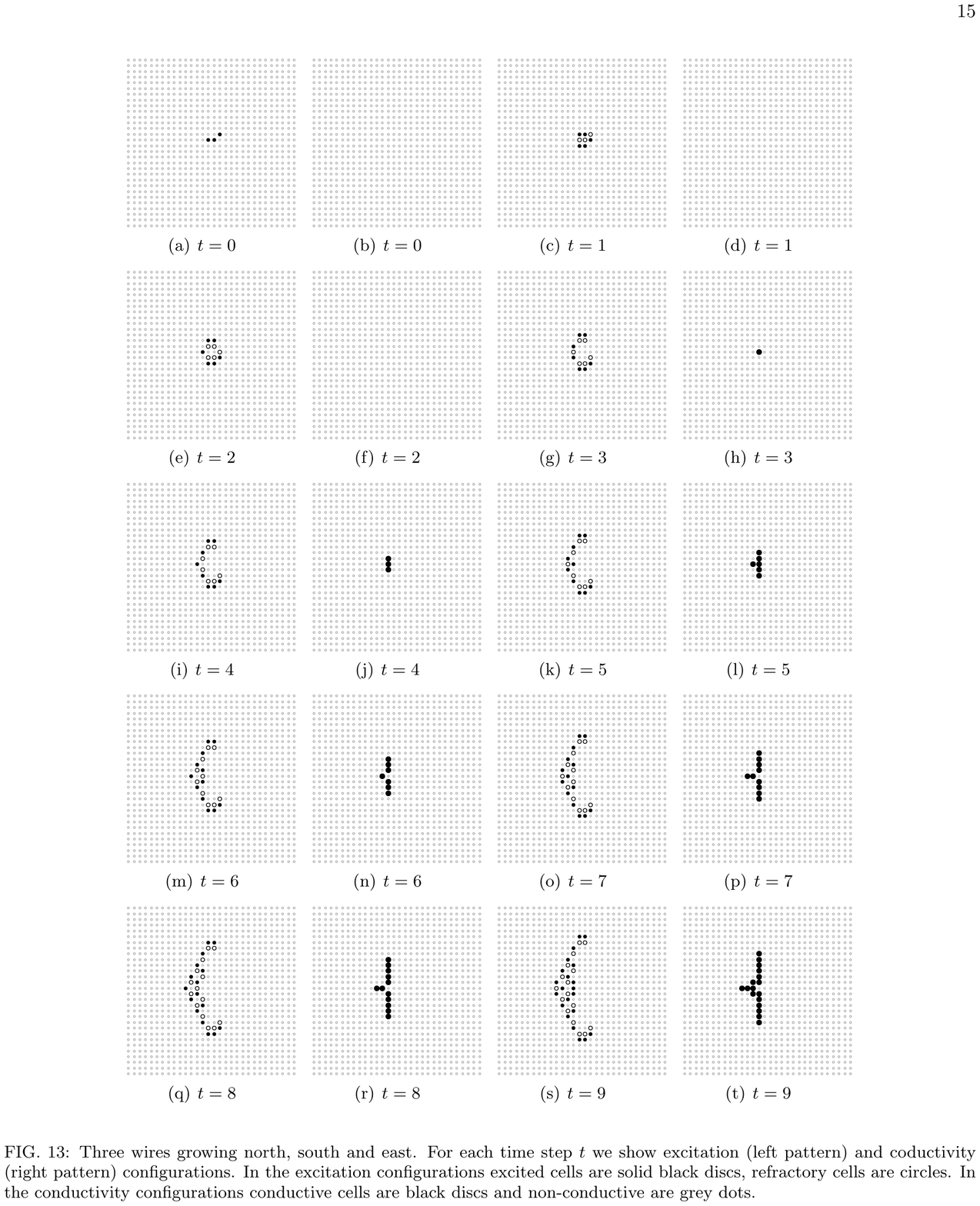}
\caption{Three wires growing north, south and east. For each time step $t$ we show excitation (left pattern) and coductivity (right pattern) configurations. In the excitation configurations excited cells are solid black discs, refractory cells are circles. In the conductivity configurations conductive cells are black discs and non-conductive are grey dots.}
\label{ThreeGrowth}
\end{figure}

To produce three wires growing north, south and east we excite a neighbouring cell north-west of the 
seed  $\begin{smallmatrix}   + & + &  \end{smallmatrix}$ (Fig.~\ref{ThreeGrowth}). The configuration
$\begin{smallmatrix}   \circ & \circ & + \\ + & + & \circ \end{smallmatrix}$ of initial excitation 
generates a wave-fragment of excited and refractory states which travels west and expands north and south
(Fig.~\ref{ThreeGrowth}ikmo). A wire growing west is produced by decreasing $\theta_1$ of excited cell
with neighbourhood 
$
\begin{smallmatrix} 
+ & \circ & - \\
- & + & \circ \\
+ & \circ & -
\end{smallmatrix}
$ 
because it has more refractory neighbours (three cells) than excited neighbours (two cells). Wires growing
north and south are produced by excited cells having neighbourhood configurations 
$
\begin{smallmatrix} 
\circ & \circ & - \\
\circ & + & \circ \\
\circ & - & \circ 
\end{smallmatrix}
$ 
and 
$
\begin{smallmatrix} 
\circ & - & \circ \\
\circ & + & \circ \\
\circ & \circ & - 
\end{smallmatrix}
$.

%\begin{figure}
%\subfigure[$t=0$]{\input{figs/SouthGrowth/0E.tex}}
%\subfigure[$t=0$]{\input{figs/SouthGrowth/0.tex}}
%\subfigure[$t=1$]{\input{figs/SouthGrowth/1E.tex}}
%\subfigure[$t=1$]{\input{figs/SouthGrowth/1.tex}}
%\subfigure[$t=2$]{\input{figs/SouthGrowth/2E.tex}}
%\subfigure[$t=2$]{\input{figs/SouthGrowth/2.tex}}
%\subfigure[$t=3$]{\input{figs/SouthGrowth/3E.tex}}
%\subfigure[$t=3$]{\input{figs/SouthGrowth/3.tex}}
%\subfigure[$t=4$]{\input{figs/SouthGrowth/4E.tex}}
%\subfigure[$t=4$]{\input{figs/SouthGrowth/4.tex}}
%\subfigure[$t=5$]{\input{figs/SouthGrowth/5E.tex}}
%\subfigure[$t=5$]{\input{figs/SouthGrowth/5.tex}}
%\subfigure[$t=6$]{\input{figs/SouthGrowth/6E.tex}}
%\subfigure[$t=6$]{\input{figs/SouthGrowth/6.tex}}
%\subfigure[$t=7$]{\input{figs/SouthGrowth/7E.tex}}
%\subfigure[$t=7$]{\input{figs/SouthGrowth/7.tex}}
%\subfigure[$t=8$]{\input{figs/SouthGrowth/8E.tex}}
%\subfigure[$t=8$]{\input{figs/SouthGrowth/8.tex}}
%\subfigure[$t=9$]{\input{figs/SouthGrowth/9E.tex}}
%\subfigure[$t=9$]{\input{figs/SouthGrowth/9.tex}}
%\caption{XXXXXX see $\begin{array}{ccc}  \- & \circ  \\ + & + & \circ \end{array}$ XXX}
%\label{SouthGrowth}
%\end{figure}

\begin{figure}[!tbp]
\includegraphics[width=\textwidth]{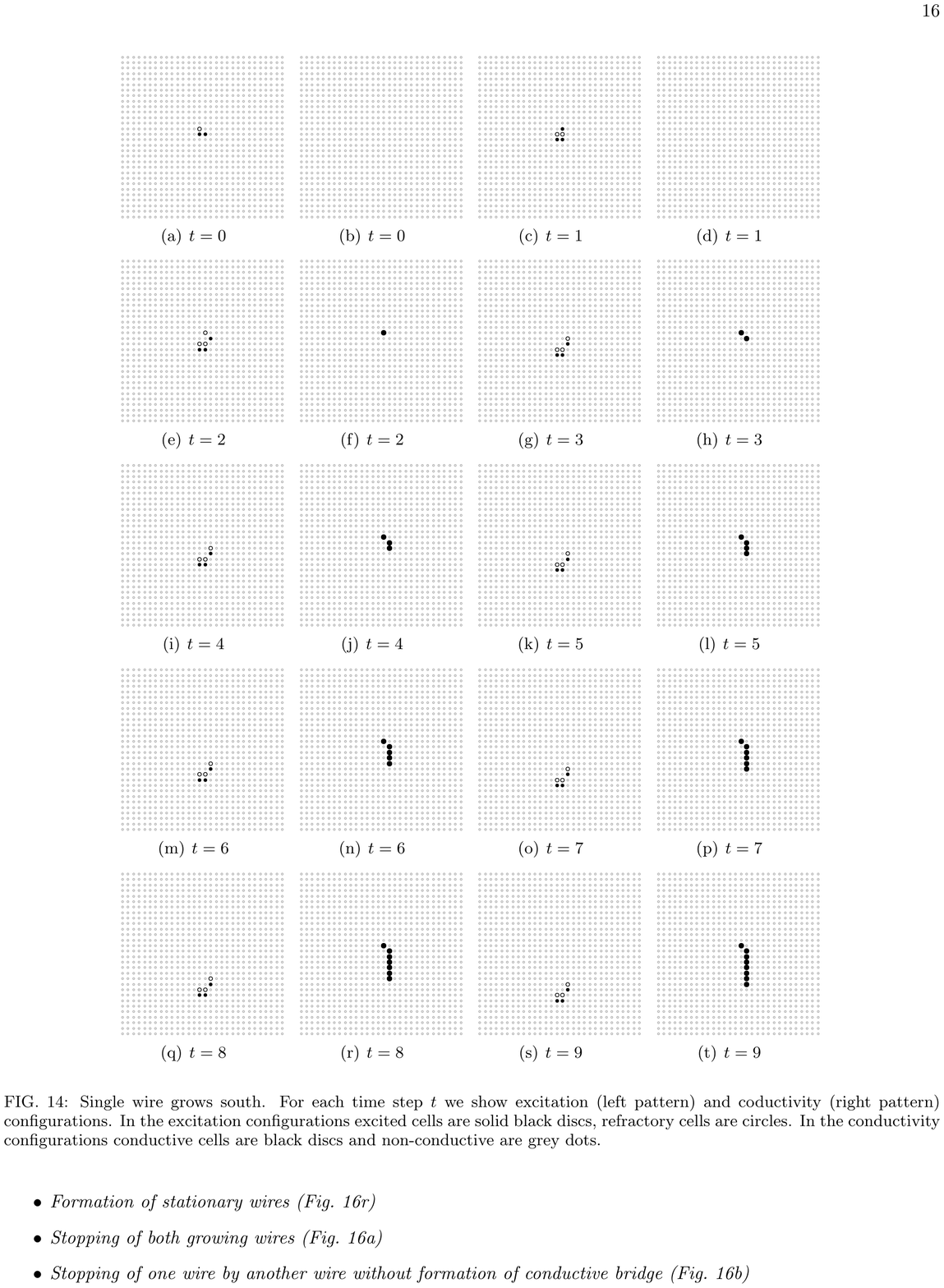}
\caption{Single wire grows south. For each time step $t$ we show excitation (left pattern) and coductivity (right pattern) configurations. In the excitation configurations excited cells are solid black discs, refractory cells are circles. In the conductivity configurations conductive cells are black discs and non-conductive are grey dots.}
\label{SouthGrowth}
\end{figure}

By making north neighbour of eastern excited cell of the seed $\begin{smallmatrix}  + & +   \end{smallmatrix}$
refractory we produce a single chain of conductive cells growing south (Fig.~\ref{SouthGrowth}).  The pattern
 $\begin{smallmatrix}  - & \circ  \\ + & +   \end{smallmatrix}$  (Fig.~\ref{SouthGrowth}a) is transformed 
 into a particle of three excited and three refractory states: 
  $\begin{smallmatrix}  
  \circ & \circ  & - \\
  \circ & \circ  & + \\
  -      & -       & \circ \\
  +      & +       & \circ \\
  \end{smallmatrix}$ 
  (Fig.~\ref{SouthGrowth}a)ikmoqs). The particle travels south. 
  Its tail leaves a trace of conductive cells  (Fig.~\ref{SouthGrowth}a)ikmoqs)jlnprt) due to
  excited cell having two refractory neighbours: 
  $\begin{smallmatrix}  
  \circ & -  & \circ \\
  \circ & +  & \circ \\
  -      & \circ      & \circ \\
  \end{smallmatrix}$, and therefore decreasing its $\theta_1$ to 1.

\begin{figure}[!tbp]
\centering
%\subfigure[$C=$(80,0)]{\includegraphics[scale=0.6]{figs/schemeOfpositionsBeforeCollision}}
%\subfigure[$C=$(80,0)]{\includegraphics[scale=0.6]{figs/HeadOn(1000)/V40_H0}}
%\subfigure[$C=$(80,1)]{\includegraphics[scale=0.6]{figs/HeadOn(1000)/V40_H1}}
%\subfigure[$C=$(80,2)]{\includegraphics[scale=0.6]{figs/HeadOn(1000)/V40_H2}}
%\subfigure[$C=$(80,3)]{\includegraphics[scale=0.6]{figs/HeadOn(1000)/V40_H3}}
%\subfigure[$C=$(80,4)]{\includegraphics[scale=0.6]{figs/HeadOn(1000)/V40_H4}}
%\subfigure[$C=$(80,5)]{\includegraphics[scale=0.6]{figs/HeadOn(1000)/V40_H5}}
%\subfigure[$C=$(81,0)]{\includegraphics[scale=0.6]{figs/HeadOn(1000)/V41_H0}}
%\subfigure[$C=$(81,1)]{\includegraphics[scale=0.6]{figs/HeadOn(1000)/V41_H1}}
%\subfigure[$C=$(81,2)]{\includegraphics[scale=0.6]{figs/HeadOn(1000)/V41_H2}}
%\subfigure[$C=$(81,3)]{\includegraphics[scale=0.6]{figs/HeadOn(1000)/V41_H3}}
%\subfigure[$C=$(81,4)]{\includegraphics[scale=0.6]{figs/HeadOn(1000)/V41_H4}}
%\subfigure[$C=$(81,5)]{\includegraphics[scale=0.6]{figs/HeadOn(1000)/V41_H5}}
\includegraphics[width=\textwidth]{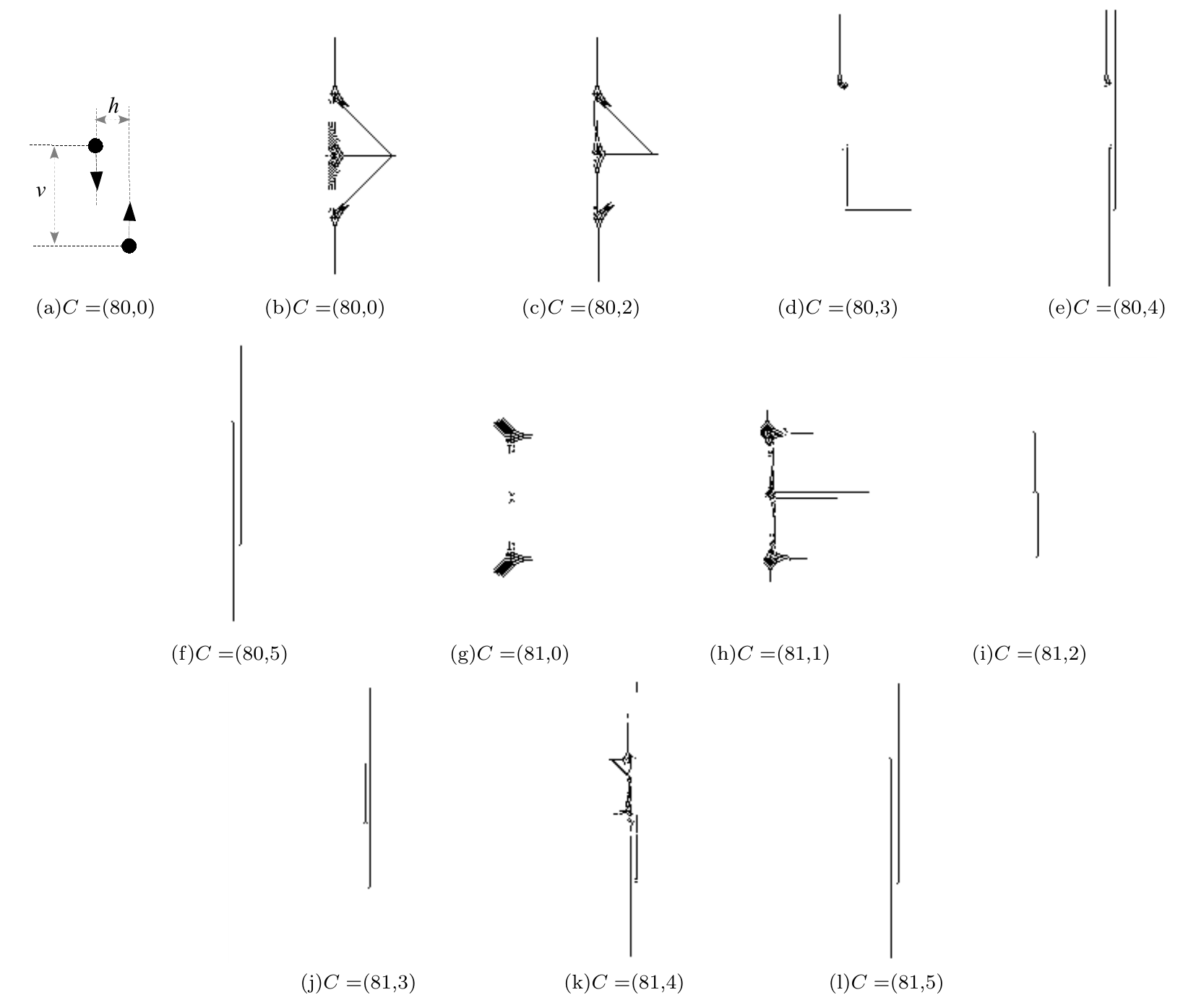}
\caption{Head on collision between wires growing north and south for various distances 
$C=(h,v)$ between their seeds. $h$ is a number of cells between westmost excited cells of seeds, 
and $v$ is a number of cells between northmost excited cells of seeds. Snapshots of configurations of 
conductive states are taken at 270th step of iterations. The outcomes of collisions are the same for
automata governed by functions $E(1,0,0,0)$ and $E(1,-1,0,0)$.} 
\label{HeadOn(1000)}
\end{figure}

\begin{figure}[!tbp]
\centering
%\subfigure[]{\includegraphics[scale=0.3]{figs/SchemeWires/V20_H10.pdf}} \hspace{0.5cm}%a
%\subfigure[]{\includegraphics[scale=0.3]{figs/SchemeWires/V20_H12.pdf}}\hspace{0.5cm}%b
%\subfigure[]{\includegraphics[scale=0.3]{figs/SchemeWires/V20_H13.pdf}}\hspace{0.5cm}%c 
%\subfigure[]{\includegraphics[scale=0.3]{figs/SchemeWires/V20_H14.pdf}}\hspace{0.5cm}%d
%\subfigure[]{\includegraphics[scale=0.3]{figs/SchemeWires/V20_H15.pdf}}\hspace{0.5cm}%e
%\subfigure[]{\includegraphics[scale=0.3]{figs/SchemeWires/V20_H17.pdf}}\hspace{0.5cm}%f
%\subfigure[]{\includegraphics[scale=0.3]{figs/SchemeWires/V20_H18.pdf}}\hspace{0.5cm}%g
%\subfigure[]{\includegraphics[scale=0.3]{figs/SchemeWires/V20_H20.pdf}}\hspace{0.5cm}%h
%\subfigure[]{\includegraphics[scale=0.3]{figs/SchemeWires/V20_H22.pdf}}\hspace{0.5cm}%i
%\subfigure[]{\includegraphics[scale=0.3]{figs/SchemeWires/V20_H23.pdf}}\hspace{0.5cm}%j
%\subfigure[]{\includegraphics[scale=0.3]{figs/SchemeWires/V40_H0.pdf}}\hspace{0.5cm}%k
%\subfigure[]{\includegraphics[scale=0.3]{figs/SchemeWires/V40_H1.pdf}}\hspace{0.5cm}%l
%\subfigure[]{\includegraphics[scale=0.3]{figs/SchemeWires/V40_H2.pdf}}\hspace{0.5cm}%m
%\subfigure[]{\includegraphics[scale=0.3]{figs/SchemeWires/V40_H3.pdf}}\hspace{0.5cm}%n
%\subfigure[]{\includegraphics[scale=0.3]{figs/SchemeWires/V40_H4.pdf}}\hspace{0.5cm}%o
%\subfigure[]{\includegraphics[scale=0.3]{figs/SchemeWires/V41_H0.pdf}}\hspace{0.5cm}%p
%\subfigure[]{\includegraphics[scale=0.3]{figs/SchemeWires/V41_H1.pdf}}\hspace{0.5cm}%q
%\subfigure[]{\includegraphics[scale=0.3]{figs/SchemeWires/V41_H2.pdf}}\hspace{0.5cm}%r
\includegraphics[width=\textwidth]{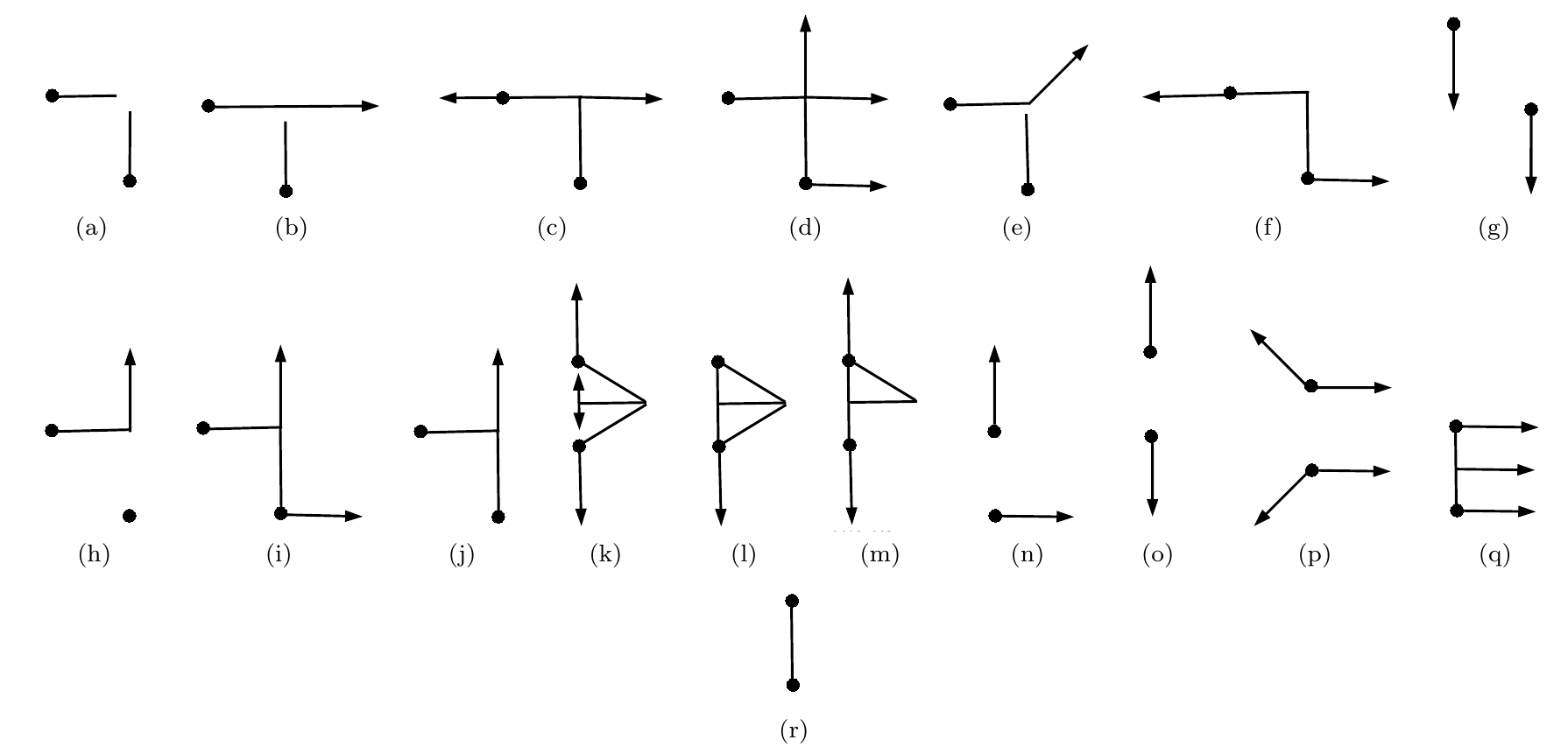}
\caption{Schemes of interactions between wires. Initial positions of seeds are shown by black discs. 
Non-growing wires are shown by line segments and growing wires by arrows. (a--j)~side collision, (k--r)~head on collision.}
\label{schemewires}
\end{figure}

We have undertook an exhaustive search on head on collisions between wires growing north and south
(Fig.~\ref{HeadOn(1000)}). A wire growing south was generated by seed
 $\begin{smallmatrix}  - & \circ  \\ + & +   \end{smallmatrix}$
 and a wire growing north by seed
  $\begin{smallmatrix}    + & + \\ - & \circ  \end{smallmatrix}$. At the beginning of each run seeds were placed at 
  distance $C=(h,v)$ from each other, where $h$ is a number of cells along rows and $v$ along columns of cellular 
  array (Fig.~\ref{HeadOn(1000)}a).  Types of collision outcomes discovered are also illustrated by schemes in  
  in Fig.~\ref{schemewires}k--r. The following basic types are found. 
  
  For distance $C=(80,0)$ the wires collide, reflect and then retract back to their origin sites. After retraction the continuing  growing directions opposite to their original velocity vectors. Wire originally growing north grows south, 
  and wire growing south grows north  (Fig.~\ref{HeadOn(1000)}b). In the same time an additional growth seed is formed 
  at the site of the collision between wires, it gives rise to two more wires growing north and south  
  (Figs.~\ref{HeadOn(1000)}b and ~\ref{schemewires}k).
  
  In situation of one-cell horizontal shift between seeds' positions, distance $C=(80,1)$, both wires reflect but 
  only the north growing wire (reflected south) continue growing beyond position of its original seed 
  (Figs.~\ref{HeadOn(1000)}c and ~\ref{schemewires}l). If there is a two cell horizontal space between the seeds 
   (Figs.~\ref{HeadOn(1000)}d) then both wires reflect and continue their growth into reflected directions
     (Figs.~\ref{HeadOn(1000)}d and ~\ref{schemewires}m). 
     
     For distance $C=(80,3)$ both wires reflect as in previous situations, however only one wire continues 
     growing north (after passing initial position of a seed) but second wire makes 90$^o$ turn when entering 
     initial position of the its seed and then grows east (Figs.~\ref{HeadOn(1000)}e and ~\ref{schemewires}n).
     In situations $C=(80,4)$ both wires just reflect without forming any conductive bridges or 
     patterns growing from the site of their collision (Figs.~\ref{HeadOn(1000)}f and ~\ref{schemewires}o).
     In situations $C=(80,5)$ and $C=(81,5)$ wires pass near each other without interacting 
     (Figs.~\ref{HeadOn(1000)}fm). 
     
     When distance between seeds is $C=(81,0)$ the growing wires reflect and detract back to their points of origination. 
     However when they reach positions of their seeds a new pattern is formed there (Figs.~\ref{HeadOn(1000)}h).  
     It exhibits multi-thread wires growing north-west, south and east from the position of northern seed, and 
     south-west, north and east from the position of southern seed (Fig.~\ref{schemewires}p).
     
     An E-shaped configuration of three wires growing east is formed when $C=(81,1)$; the seeds' site are also connected by a conductive wire  (Figs.~\ref{HeadOn(1000)}i and ~\ref{schemewires}q).
     
     \begin{finding}
By positioning seeds $\begin{smallmatrix} + & + \\ - & \circ \end{smallmatrix}$ and 
$\begin{smallmatrix} -  & \circ \\ + & +  \end{smallmatrix}$ with shift $C=(odd, 2)$ it is possible to make a stationary 
conductive wire between the seeds' locations. 
\end{finding}
     
     For example, in situation $C=(81,2)$ (Figs.~\ref{HeadOn(1000)}j) the growing wires collide, make a conductive bridge and stop their propagation. Thus a stationary wire is formed connecting sites of southern and northern seeds  (Fig.~\ref{schemewires}r). The seeds' sites are connected by a wire in $C=(81,3)$  (Figs.~\ref{HeadOn(1000)}k) and
     $C=(81,4)$ (Figs.~\ref{HeadOn(1000)}l) however the wire continue expanding north and south.

%\begin{figure}
%\centering
%\subfigure[(40,0)]{\includegraphics[scale=0.5]{figs/HeadOn(1-100)/V40_H0_(1-100)}}
%\subfigure[(40,1)]{\includegraphics[scale=0.5]{figs/HeadOn(1-100)/V40_H1_(1-100)}}
%\subfigure[(40,2)]{\includegraphics[scale=0.5]{figs/HeadOn(1-100)/V40_H2_(1-100)}}
%\subfigure[(40,3)]{\includegraphics[scale=0.5]{figs/HeadOn(1-100)/V40_H3_(1-100)}}
%\subfigure[(40,4)]{\includegraphics[scale=0.5]{figs/HeadOn(1-100)/V40_H4_(1-100)}}
%\subfigure[(40,5)]{\includegraphics[scale=0.5]{figs/HeadOn(1-100)/V40_H5_(1-100)}}
%\subfigure[(41,0)]{\includegraphics[scale=0.5]{figs/HeadOn(1-100)/V41_H0_(1-100)}}
%\subfigure[(41,1)]{\includegraphics[scale=0.5]{figs/HeadOn(1-100)/V41_H1_(1-100)}}
%\subfigure[(41,2)]{\includegraphics[scale=0.5]{figs/HeadOn(1-100)/V41_H2_(1-100)}}
%\subfigure[(41,3)]{\includegraphics[scale=0.5]{figs/HeadOn(1-100)/V41_H3_(1-100)}}
%\subfigure[(41,4)]{\includegraphics[scale=0.5]{figs/HeadOn(1-100)/V41_H4_(1-100)}}
%\subfigure[(41,5)]{\includegraphics[scale=0.5]{figs/HeadOn(1-100)/V41_H5_(1-100)}}
%\caption{Head on collision...(1-100) }
%\label{HeadOn(1-100)}
%\end{figure}

\begin{figure}[!tbp]
\centering
%\subfigure[$C=$(40,0)]{\includegraphics[scale=0.6]{figs/CrossCollisionschemeOfpositionsBeforeCollision}}
%\subfigure[$C=$(60,40)]{\includegraphics[scale=0.6]{figs/CrossCollision/CrossCollisionV20_H10_(1000)}}
%%\subfigure[$C=$(60,41)]{\includegraphics[scale=0.6]{figs/CrossCollision/CrossCollisionV20_H11_(1000)}}
%\subfigure[$C=$(60,42)]{\includegraphics[scale=0.6]{figs/CrossCollision/CrossCollisionV20_H12_(1000)}}
%\subfigure[$C=$(60,43)]{\includegraphics[scale=0.6]{figs/CrossCollision/CrossCollisionV20_H13_(1000)}}
%\subfigure[$C=$(60,44)]{\includegraphics[scale=0.6]{figs/CrossCollision/CrossCollisionV20_H14_(1000)}}
%\subfigure[$C=$(60,45)]{\includegraphics[scale=0.6]{figs/CrossCollision/CrossCollisionV20_H15_(1000)}}
%\subfigure[$C=$(61,40)]{\includegraphics[scale=0.6]{figs/CrossCollision/CrossCollisionV20_H16_(1000)}}
%\subfigure[$C=$(61,41)]{\includegraphics[scale=0.6]{figs/CrossCollision/CrossCollisionV20_H17_(1000)}}
%\subfigure[$C=$(61,42)]{\includegraphics[scale=0.6]{figs/CrossCollision/CrossCollisionV20_H18_(1000)}}
%\subfigure[$C=$(61,43)]{\includegraphics[scale=0.6]{figs/CrossCollision/CrossCollisionV20_H19_(1000)}}
%\subfigure[$C=$(61,44)]{\includegraphics[scale=0.6]{figs/CrossCollision/CrossCollisionV20_H20_(1000)}}
%\subfigure[$C=$(61,45)]{\includegraphics[scale=0.6]{figs/CrossCollision/CrossCollisionV20_H21_(1000)}}
\includegraphics[width=\textwidth]{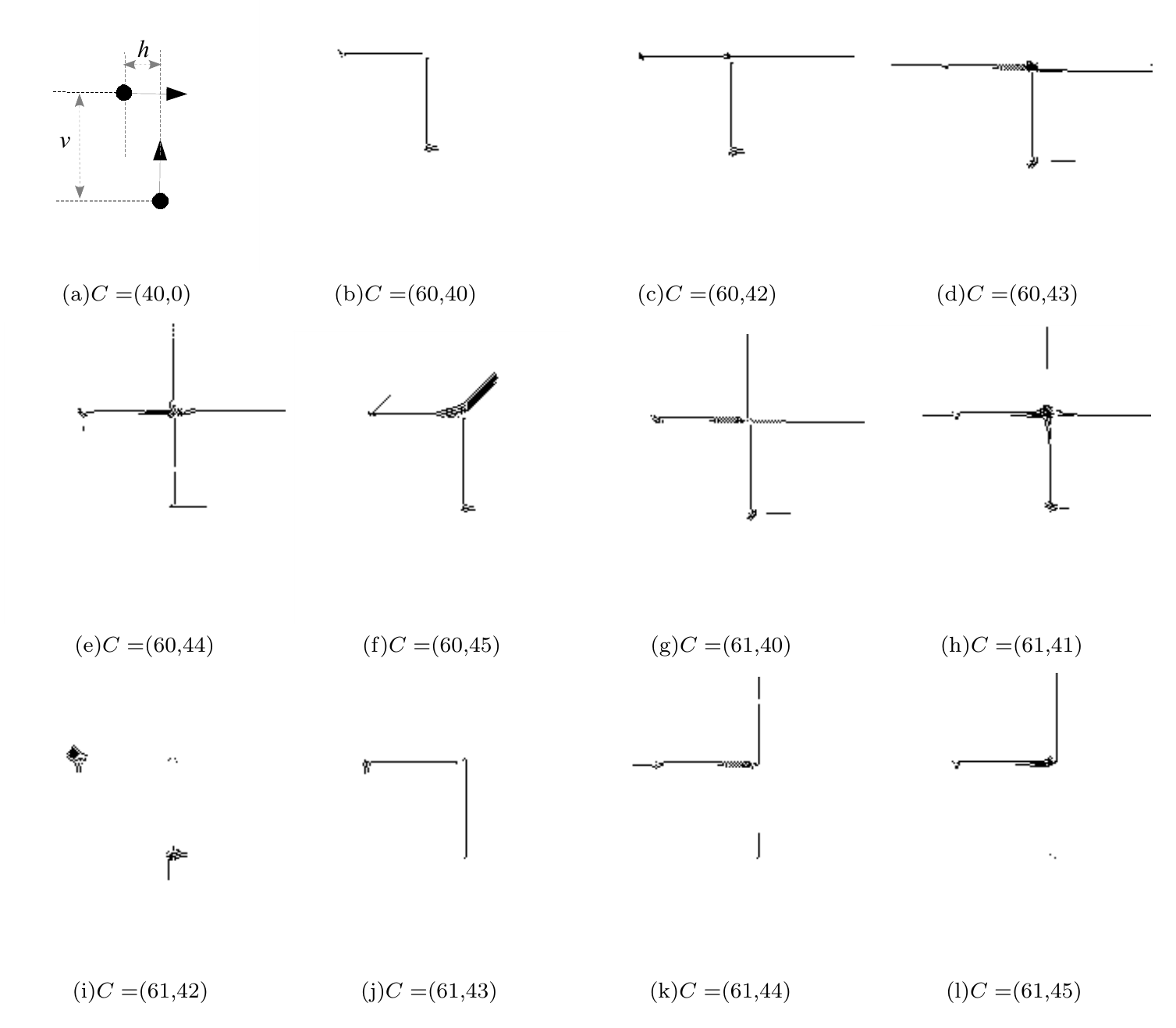}
\caption{Side collision between wires growing north and east for various distances 
$C=(h,v)$ between their seeds. $h$ is a number of cells between northmost excited cells of seeds, 
and $v$ is a number of cells between northmost excited cells of seeds. Snapshots of configurations of 
conductive states are taken at 300th step of iterations. The outcomes of collisions are the same for
automata governed by functions $E(1,0,0,0)$ and $E(1,-1,0,0)$.}
\label{sidecollision}
\end{figure}

To study outcomes of side collisions between wires we position two seeds
$
\begin{smallmatrix}    + & + \\ - & \circ  \end{smallmatrix}
$
and 
$
\begin{smallmatrix}    -        & + \\ 
                                 \circ  & +  \end{smallmatrix}
$
at distance $C=(h,v)$ (Fig.~\ref{sidecollision}a), where $h$ is a number of cells between northmost excited cells of seeds, 
and $v$ is a number of cells between northmost excited cells of seeds. The seed 
$
\begin{smallmatrix}    + & + \\ 
                                 - & \circ  \end{smallmatrix}
$
leads to formation of a single thread wire growing north. The seed
$
\begin{smallmatrix}    -        & + \\ 
                                 \circ  & +  \end{smallmatrix}
$
generates a wire east. Configurations of cells in conductive states developed on 300th step after excitation of 
automaton with the seeds are shown in Fig.~\ref{sidecollision}. 

Colliding wires stop short of touching each other and do not propagate anymore for $C=(60,40)$ and 
$C=(61,43)$ (Figs.~\ref{sidecollision}bk and~\ref{schemewires}a). Growth of wire propagating north is cancelled by 
wire propagating east in $C=(60,42)$ (Fig.~\ref{schemewires}b), the wires do contact each other  (Figs.~\ref{sidecollision}c).

In condition $C=(60,43)$  (Figs.~\ref{sidecollision}d) two new growing wires are formed in the result of collision 
of a wire travelling north to a wire travelling east. One new wire propagates west and another wire propagates east (Fig.~\ref{schemewires}c).

Three new growing wires are formed in collision of north and east propagating wires when distance between their
seeds is $C=(60,44)$  (Figs.~\ref{sidecollision}e and~\ref{schemewires}d) and $C=(61,40)$ 
 (Figs.~\ref{sidecollision}g and~\ref{schemewires}d). One new wire propagates north and two new wires propagate east.  The south wire propagating east is formed when originally north propagating wire retracts in 
the result of collision with originally east propagating wire and is reflected by configurations of excitation interval boundaries its imposed by itself. 

Almost elastic like reflection of a wire is observed in case $C=(60,45)$ 
(Figs.~\ref{sidecollision}f and~\ref{schemewires}e).  A wire growing north collides and got cancelled by a wire growing east. In the result of impact the east growing wire is reflected and starts growing north-east 
(Fig.~\ref{schemewires}e). 

In situation $C=(61,42)$ both wires retract in the result of collision. However, when they reach sites of their origination (where cells have already updated boundaries of excitation interval) they are transformed into extended patterns which give rise to two new wires, both growing south (Figs.~\ref{sidecollision}g and~\ref{schemewires}j). 

In situation $C=(61,44)$ and $C=(61,45)$ a wire growing north is retracted back to its seed's position in the result of collision with a wire growing east (Figs.~\ref{sidecollision}h and~\ref{schemewires}h). In the same time the wire growing east is reflected and turns north.

\begin{finding}
It is possible to implement universal routing of conductive wires by positioning seeds of growing wires, the following operations with wires are implementable:
\begin{itemize}
\item Formation of stationary wires (Fig.~\ref{schemewires}r)
\item Stopping of both growing wires (Fig.~\ref{schemewires}a)
\item  Stopping of one wire by another wire without formation of conductive bridge (Fig.~\ref{schemewires}b)
\item  Formation of conductive circuit with one growing wire (Fig.~\ref{schemewires}l)
\item  Formation of conductive circuit with two growing wires (Fig.~\ref{schemewires}cfikm)
 \item Formation of conductive circuit with three growing wires (Fig.~\ref{schemewires}dq)
 \item Reflection of wires without conductive bridging  (Fig.~\ref{schemewires}hno)
 \item Stopping of one wire and reflection of another  (Fig.~\ref{schemewires}e)
\item Co-orientation of both growing wires without formation of conductive bridge (Fig.~\ref{schemewires}g)
\item Symmetric reflection and multiplication without formation conductive bridging (Fig.~\ref{schemewires}p).  
\end{itemize}
\end{finding}

\section{Summary}
\label{discussion}

We introduced a two-dimensional excitable cellular automaton where resting cells excite depending on 
whether numbers of their excited neighbours belong to excitation intervals and boundaries of the excitation 
intervals are updated depending on ratio of excited and refractory cells in each cell's neighbourhood. We defined 
conductivity of a cell via size of its excitation interval and selected the excitation interval update functions that lead
to formation of connected configurations of conductive cells. 

We demonstrated that by positioning elementary seeds  of excitation we grow conductive wires (chains of cells in conductive states) and implement routing of the wires via collisions between the wires. Results presented might shade a light onto development of information pathways in excitable spatially extended media and contribute towards manufacturing of self-growing and self-organising circuits in ensembles of organic memristive polymers. 

Principle findings of the paper are following: 
\begin{itemize}
\item We demonstrated that it is possible to fine tune conductivity of an excitable medium by controlling local dynamics of excitation.
\item Functions which stabilise excitation dynamics (where size of excitation interval increase with decrease of excitation and decreases when excitation dominates) generate fully conductive when a small number of initially resting cells are stimulated.
\item A point-wise initial excitation can play a seed of a growing wire, a chain of cells in conductive states; directions of the wire grows in pre-programmed in the configuration of initial excitation
\item The growing wires can be routed in almost arbitrary manner, dependent on positions of their seeds
\item Several wires can interact with each other by changing directions of their growth, merging in a single wire and co-aligning.
\end{itemize}

We shown how to design and grow potential information pathways however we did not study how the information can be processed in the conductive configurations and circuits. In many cases of extended patterns are formed at the sites of collision between growing wires.  Chances are high that these patterns can implement can implement a range of sensible transformations input excitation to output excitation, which could be interpreted in terms of computation. Computational abilities of the conductive circuits grown in excitable cellular automata will be a major topic of further studies.

\end{document}